\documentclass[aps,twocolumn,nofootinbib,preprintnumbers,superscriptaddress]{revtex4-2}

\usepackage{color}
\usepackage{bbm}
\usepackage{tikz}
\usepackage{braket}
\usepackage{comment}
\usepackage{physics}

\usepackage[
pagebackref=false,
colorlinks=true,
linkcolor=blue,
urlcolor=blue,
filecolor=black,
citecolor=red,
pdfstartview=FitV,
pdftitle={},
pdfauthor={},
pdfsubject={},
pdfkeywords={},
pdfpagemode=None,
bookmarksopen=true
]{hyperref}

\usepackage[normalem]{ulem}
\usepackage{amsmath, amssymb, bm}
\usepackage{enumerate}
\usepackage{amsfonts}
\usepackage{epsfig}
\usepackage{mathbbol}
\usepackage{mathrsfs}
\definecolor{darkgreen}{rgb}{0.0, 0.5, 0.0}
\usepackage{subcaption}
\usepackage{braket}
\usepackage{bm}
\newcommand{\be}{\begin{equation}}
\newcommand{\ee}{\end{equation}}
\newcommand{\f}{\frac}
\newcommand{\s}{\sqrt}

\def\ba#1\ea{\begin{align}#1\end{align}}

\usepackage{xcolor}

\begin{document}

\title{Dynamical quantum phase transitions in SYK Lindbladians}

\author{Kohei Kawabata}
\thanks{The authors are listed in alphabetical order.}
\affiliation{Department of Physics, Princeton University, Princeton, New Jersey 08544, USA}
\affiliation{Institute for Solid State Physics, University of Tokyo, Kashiwa, Chiba 277-8581, Japan}

\author{Anish Kulkarni}
\thanks{The authors are listed in alphabetical order.}
\affiliation{Department of Physics, Princeton University, Princeton, New Jersey 08544, USA}

\author{Jiachen Li}
\thanks{The authors are listed in alphabetical order.}
\affiliation{Department of Physics, Princeton University, Princeton, New Jersey 08544, USA}

\author{Tokiro Numasawa}
\thanks{The authors are listed in alphabetical order.}
\affiliation{Institute for Solid State Physics, University of Tokyo, Kashiwa, Chiba 277-8581, Japan}

\author{Shinsei Ryu}
\thanks{The authors are listed in alphabetical order.}
\affiliation{Department of Physics, Princeton University, Princeton, New Jersey 08544, USA}

\date{\today}

\begin{abstract}
We study the %Lindbladian 
open quantum
dynamics of the Sachdev-Ye-Kitaev (SYK) model 
described by the Lindblad master equation, 
where the SYK model is coupled to Markovian reservoirs
with jump operators that are either linear or quadratic in the
Majorana fermion operators. 
%Here, the linear jump operators are non-random while the quadratic jump operators are sampled from a Gaussian distribution. 
Of particular interest for us is 
the time evolution of
the %so-called 
dissipative form factor, 
%the disorder averaged trace of the exponentiated Lindbladian. 
which quantifies the average overlap between the initial and time-evolved density matrices as an open quantum generalization of the Loschmidt echo.
We %found 
find
that the dissipative form factor
exhibits %various 
dynamical 
quantum
phase transitions.
%many of which are unique to open quantum many-body systems.
%For the %case of 
%linear non-random jump operators, 
We %found 
analytically demonstrate
a 
discontinuous 
dynamical phase transition %akin to
in the limit of large number of fermion flavors,
which is formally akin to
the %finite temperature 
thermal phase
transition in
the two-coupled SYK model between 
the black-hole and wormhole phases.
%In addition, we also found 
%We also find continuous dynamical
%phase transitions that are unique to
%open quantum many-body systems. 
%\textcolor{red}{[Actually I'm not sure --
%this is rather vague. 
%It's unique in the sense that 
%the 1st order phase transition is 
%analogous to the Hawking-Page transition
%for the hermitian model,
%but continous transitions do not 
%have such counter parts. 
%It would be good if we can say more about 
%this point. 
%Maybe we can say
%"
%We also find continuous dynamical
%phase transitions that 
%do not have counterparts 
%in 
%the hermitian model.
%"
%}
We also find continuous dynamical phase transitions that do not have counterparts in the 
%original 
two-coupled SYK model.
While the phase transitions are sharp in the limit of large number of fermion flavors,
their qualitative signatures are present 
even for the finite number of fermion flavors,
as we show numerically.
\end{abstract}

\maketitle

%\tableofcontents

\section{Introduction}

%Open quantum systems exhibit many phenomena which do not occur
%in closed systems governed by unitary dynamics.
%For example,
%quantum phase transitions can be
%induced by introducing dissipation, or by measuring
%(monitoring) quantum systems.
%Topological states of matter can also be prepared
%by engineering dissipative dynamics. 
%Non-hermiticity can also lead to
%topological phenomena that do not have counter parts
%in Hermitian systems. 
%The complex interplay of strong interactions
%and dissipation however has not been systematically explored.

The physics of open quantum systems has recently attracted growing interest.
Since 
%the 
coupling to the external environment is unavoidable in realistic physical systems, an understanding of open quantum systems is important for quantum technology~\cite{Breuer-textbook}.
Notably, dissipation is not necessarily a nuisance that destroys quantum coherence and the concomitant quantum phenomena; 
rather, dissipation can even lead to new physical phenomena that have no analogs in closed quantum systems.
For example, engineered dissipation can be utilized to prepare a desired quantum state~\cite{Verstraete-09, Diehl-08, Diehl-11}.
Dissipation can also give rise to unique non-Hermitian topological phenomena~\cite{Bergholtz-review}.
Furthermore, open quantum systems exhibit phase transitions that cannot occur in closed quantum systems at thermal equilibrium~\cite{Yang-Lee-52I, *Lee-Yang-52II, Fisher-78, Caldeira-Leggett-PRL81, *Caldeira-Leggett-AP83, *Leggett-review, Bender-02, Albert-14, Lee-14, Minganti-18, Dora-19, Hayata-21}.
Prime recent examples include the entanglement phase transitions induced by the competition between the unitary dynamics and the quantum measurements~\cite{Chan-19, Skinner-19, Li-18, *Li-19, Choi-20, Gullans-20}. 
Despite these recent advances, the interplay of strong many-body interactions and dissipation, as well as the consequent phase transitions, has yet to be fully understood.

%In \cite{kulkarni2021syk, S__2022},
%to better understand the effect of dissipation in strongly coupled systems,
%the Sachdev-Ye-Kitaev model in the presence of dissipation was introduced
%and studied.
%This is a nice toy model that allows many exact analyses
%in the limit of a large number of fermion flavors.
%Furthermore, the model exhibits 
%among others, a transition between coherent and overdamped dynamics
%as we change the strength of dissipation.
%Due to the competition between strong correlations and dissipation,
%the decay rate exhibits non-monotonic behavior as a function of disorder strength.

In the theory of phase transitions, it is important to develop a prototypical model that captures the universal behavior. 
Recently, open quantum generalizations of the Sachdev-Ye-Kitaev (SYK) model~\cite{Sachdev-Ye-93, Kitaev-KITP15} were proposed in Refs.~\cite{S__2022, kulkarni2021syk} as a prototype of open quantum many-body systems.
In this model, dissipation is formulated by the Lindblad master equation~\cite{GKS-76, Lindblad-76}, which is different from the non-Hermitian SYK Hamiltonians~\cite{ChunxiaoLiu-21, GarciaGarcia-22PRL, PengfeiZhang-21, *Jian-21, GarciaGarcia-22PRX}.
The original SYK Hamiltonian is a fermionic model with 
%the 
fully-coupled random interactions and exhibits 
%the 
quantum chaotic behavior~\cite{Sachdev-Ye-93, Kitaev-KITP15, Polchinski-Rosenhaus-16, Maldacena-Stanford-16, Gu-17, Cotler-17, Song-17, Rosenhaus-review, Sachdev-review}.
Similarly, the SYK Lindbladian is a prototype that exhibits the strongly-correlated chaotic behavior of open quantum systems~\cite{Grobe-88, Xu-19, Hamazaki-19, Denisov-19, Can-19PRL, *Can-19JPhysA, Hamazaki-20, Akemann-19, Sa-20, Wang-20, Xu-21, *Cornelius-22, JiachenLi-21}.
As an advantage, the SYK Lindbladian is analytically tractable in the limit of the large number of fermion flavors even in the presence of dissipation. 
In Refs.~\cite{S__2022, kulkarni2021syk}, the decay rate was analytically calculated in this limit, by which a transition between the underdamped and overdamped regimes was demonstrated.
Still, the open quantum dynamics of the SYK Lindbladians remains mainly unexplored.
As a prototype of open quantum many-body systems, the investigation into the SYK Lindbladians should deepen our general understanding of open quantum physics.

%In this paper, we
%further study this model
%by looking at 
%the real-time dynamics of this model and explore dynamical phase transitions.
%Specifically, we study the dissipative form factor defined below.

In this work, we find the dynamical quantum phase transitions in the SYK Lindbladians.
We study the open quantum dynamics of the SYK Lindbladians and especially focus on the time evolution of the dissipative form factor.
This quantifies the average overlap between the initial and time-evolved density matrices and serves as a partition function of the open quantum dynamics, similarly to the Loschmidt echo for the unitary dynamics of closed quantum systems.
We find the singularities of the dissipative form factor as a function of time, which signal the dynamical quantum phase transitions similarly to the unitary counterparts~\cite{Heyl-13, Heyl-14, Budich-16, Heyl-15, Sharma-16, Flaschner-18, Jurcevic-17, Zhang-17, Hamazaki-21, Heyl-review}.
Notably, this quantum phase transition appears only in the dynamics in 
%contrasts 
contrast 
with the conventional phase transitions for thermal equilibrium or ground states.
In particular, we investigate the SYK Hamiltonian coupled to Markovian nonrandom linear dissipators and random quadratic dissipators.
In the limit of the large number $N$ of fermions, we analytically obtain the dissipative form factor and demonstrate the discontinuous dynamical phase transition, %for the non-random linear dissipators,
which is formally akin to the thermal phase transition in the two-coupled SYK model between the black-hole and wormhole phases~\cite{2018arXiv180400491M}.
%For the random quadratic dissipators, on the other hand, we 
We also show the continuous dynamical transition that has no counterparts in the original two-coupled SYK model.
Furthermore, we numerically show that signatures of the dynamical quantum phase transitions remain to appear even for finite $N$ although the singularities are not sharp.

The rest of this work is organized as follows.
In
Sec.~\ref{Lindbladians and dissipative form factor},
we start by introducing the models,
and review the quantity of our interest,
the dissipative form factor. 
In 
Sec.~\ref{Non-random linear jump operators},
we study the SYK Lindbladian 
with nonrandom linear jump operators. 
We discuss both numerics of the large 
$N$ saddle point equations
and the analytical approach 
in the large $q$ limit.
In 
Sec.~\ref{Random quadratic jump operators},
we study the SYK Lindbladian
with random quadratic jump operators.
We conclude in 
Sec.~\ref{Discussion}.

\section{SYK Lindbladians and dissipative form factor}
    \label{Lindbladians and dissipative form factor}

We consider Markovian dynamics of
the density matrix $\rho(t)$ described by
the Lindblad master equation~\cite{Breuer-textbook}:
\begin{align}
  \frac{d }{dt}
  \rho(t)
  &= \mathcal{L}(\rho(t)) \nonumber \\
  &\equiv
    - i [H,\rho(t)]
    \nonumber \\
  &\qquad 
    +
    \sum_{a}
    \left[
    L^a \rho(t) L^{a \dag}
    -
    \frac{1}{2}
    \{
    L^{a \dag} L^{a}, \rho(t)
    \}
    \right],
\end{align}
where $H$ is the Hamiltonian,
and $\{L^a\}$ is a set of jump operators that describe the dissipative process with the external environment.
%We will study two SYK Lindbladian models. In bothe models,
In our models,
the Hamiltonian is given by the $q$-body SYK Hamiltonian
\begin{align}
    \label{eq: SYK Hamiltonian}
	H^{\text{SYK}} = i^{q/2} \sum_{1 \leq i_1 < i_2 < \cdots < i_q \leq N} J_{i_1 i_2 \cdots i_q} \psi^{i_1} \psi^{i_2} \cdots \psi^{i_q}.
\end{align}
Here, $\psi^{i=1, \dots, N}$ are Majorana fermion operators satisfying $\{ \psi^i , \psi^j \} = \delta_{ij}$. $J_{i_1 \cdots i_q}$ are real independent Gaussian distributed random variables with zero mean and variance given by
\begin{align}
\overline{
	(J_{i_1 \cdots i_q})^2
	}
	 = \sigma_J^2= \frac{J^2(q-1)!}{N^{q-1}} \quad \left( J\in \mathbb R^+ \right),
\end{align}
where 
%the bracket 
$
\overline{\cdots}
$
denotes the disorder average.
We consider two choices of the jump operators:
nonrandom linear and random quadratic. 
The nonrandom linear jump operators are
\begin{equation}
\label{eq: non-random jumps}
L^i = \s{\mu}\psi^i
\quad
\left( i=1,\cdots, N,~\mu \in \mathbb{R}^+ \right).
\end{equation}
On the other hand, 
the random 
%quadratic 
$p$-body 
jump operators are
\begin{equation}
    \label{p-body jumps}
	L^a = \sum_{1\leq i_1 < \cdots < i_p \leq N} K^a_{i_1 \cdots i_p} \psi^{i_1} \cdots \psi^{i_p} \quad \left( a=1, 2,\dots , M \right),
\end{equation}
where
$K^a_{i_1 \cdots i_p}$ are complex Gaussian random variables with zero mean and variance given by
\begin{equation}
    \label{eq: random jumps}
    \overline{
	|K^a_{i_1 \cdots i_p}|^2} = \sigma_K^2 = \frac{K^2(p-1)!}{N^p} \quad \left( K\in \mathbb{R}^+ \right).
\end{equation}
In this work, we focus on $p=2$ for clarity.

Since the Lindbladian is a superoperator that acts on the density matrix, it is useful to introduce the operator-state map. 
Here,
%by the operator-state map,
we %"
vectorize %'' 
the density matrix $\rho (t)$
and regard it as a state $\ket{\rho (t)}$ in the doubled Hilbert space
$\mathcal{H}_+\otimes \mathcal{H}_-$\footnote{
Strictly speaking,
this is a sloppy notation 
%is a misnomer 
when we discuss
the operator-state map 
for fermionic systems
since 
states in the $+$ and $-$
sectors may not commute
because of the 
Fermi statistics~\cite{Dzhioev-11, *Dzhioev-12}.
}.
Correspondingly, we regard the Lindbladian
as an operator acting on the doubled Hilbert space.
%For fermionic systems,
%including the SYK-type models relevant to this work,
%we need to use the ``fermionic''
%operator-state map that properly takes
%into account the fermi statistics.
For the SYK-type models relevant to this work,
the Lindbladian acting %of 
on
the doubled Hilbert space is given by
\begin{align}
  \mathcal{L}
  &=
  -i H^+ + i (-i)^q H^-
  \nonumber \\
  &\qquad
    +
    \sum_a
    \left[
    (-i)^pL^a_+ L^{a\dagger}_-
    -
    \frac{1}{2}
    L^{a\dagger}_+ L^a_+
    -
    \frac 12 L^a_- L^{a\dagger}_-
    \right],
\end{align}
where $H^{\pm}$ and $L_{\pm}^a$ act on $\mathcal{H}_{\pm}$, respectively. 
%The Hamiltonian here is a $q$-body operator in terms of fermion operators.
While the Hamiltonian part acts only on
the individual bra or ket space, the dissipation term couples these two spaces.

A quantity of our central interest is
the disorder-averaged  
trace of the exponential of the Lindbladian,
\begin{equation}
  \label{eq: DFF definition}
  F(T_L) = 
  %\expectationvalue{
  \overline{
    \mathrm{Tr}_{\mathcal{H}_+ \otimes \mathcal{H}_-}\, (e^{T_L\mathcal L})},
    %},
\end{equation}
which we call the dissipative form factor~\cite{Can-19PRL, *Can-19JPhysA, Xu-21, *Cornelius-22}.
The trace in Eq.~\eqref{eq: DFF definition}
is taken over the doubled Hilbert space $\mathcal{H}_+ \otimes \mathcal{H}_-$.
As explained shortly, the dissipative form factor quantifies the average overlap between the initial and time-evolved density matrices and serves as the Loschmidt echo of open quantum systems.
In the following,
we obtain
the dissipative
form factor 
of the SYK Lindbladians,
using both the analytical calculations for large $N$ %large $N$ saddle points
and the numerical calculations for finite $N$,
and demonstrate its singularities in the open quantum dynamics---dynamical quantum phase transitions. %finite $N$ numerics
%for the dissipative SYK models.
More specifically, we analyze  
(a dissipative analog
of)
the rate function of
the dissipative form factor:
%the Loschmidt echo:
\begin{equation}
  \label{eq: rate function}
  i \mathcal S(T_L) %= 
  \equiv
  \lim_{N\rightarrow\infty} \frac{\log F(T_L)}{N}.
\end{equation}
Although the spectrum 
of the Lindbladian 
is complex in general,
the rate function
is always real
since the spectrum 
is symmetric about
the real axis. 
%If we remove all jump operators (i.e. set them to zero), then this definition
%coincides with the unitary analogs. In this case, non-analiticities in the rate
%function have been proposed as a diagnostic of dynamical phase transitions
%\cite{Heyl-review}.
%We wish to extend this idea to the non-unitary case.
%
%\magenta{Furthermore, $i \mathcal S$ monotonically decreases as a function of $T_{L}$ because of the contractive nature of the Lindbladian dynamics.}
We also note that the dissipative form factor does not depend on initial conditions but is determined solely by the Lindbladian.
%\magenta
{At $T_{L} = 0$, we always have $F (T_{L}=0) = 2^{N}$ and hence $i \mathcal S(T_L=0) = \log 2$.}

Several motivating comments are in order.
First, %in the limit of vanishing dissipation,
in the absence of dissipation,
the dissipative form factor in Eq.~\eqref{eq: DFF definition}
coincides with the spectral form factor of %a hermitian Hamiltonian, 
Hermitian Hamiltonians.
In fact, we have
\begin{align}
&
  \mathrm{Tr}_{\mathcal{H}_+\otimes\mathcal{H}_-}\,(e^{T_L\mathcal L})
  \nonumber \\
  &\quad=
  \mathrm{Tr}_{\mathcal{H}_+}\big(
  e^{ - i T_L H^+}
  \big)
  \, 
  \mathrm{Tr}_{\mathcal{H}_-}
  \big( e^{+ i T_L (-i)^q H^-} 
  \big)
    \nonumber \\
  &\quad=
    \left|
    \mathrm{Tr}_{\mathcal{H}}\left(
    e^{ - i T_L H}
    \right)
    \right|^2,
\end{align}
where in the last line we take %$q\equiv 4$ 
$q \equiv 0$ (mod $4$)
for simplicity.
%We would thus expect that 
%the dissipative form factor \eqref{eq: DFF definition}
%can capture some properties of the
%Lindbladian spectrum. 
Since the spectral form factor captures the quantum chaos of the SYK-type Hamiltonians~\cite{Cotler-17},
we expect that the dissipative form factor in Eq.~\eqref{eq: DFF definition} also captures the quantum chaos of the SYK Lindbladians.
Accordingly, %in the limit of vanishing jump operators,
in the absence of the dissipation,
the definition of the rate function in Eq.~\eqref{eq: rate function}
%If we remove all jump operators (i.e. set them to zero), then this definition
coincides with its unitary analog. 
%In this case, 
In the unitary case,
%non-analiticities 
non-analytical behavior
in the rate
function 
as a function of time
%have been 
was
proposed as a diagnostic of the dynamical quantum phase transitions~\cite{Heyl-13, Heyl-14, Budich-16, Heyl-15, Sharma-16, Flaschner-18, Jurcevic-17, Zhang-17, Hamazaki-21, Heyl-review}.
Here, we %wish to 
extend this idea to the non-unitary case.
We also note that there %are other similar quantities,
is another related quantity,
dissipative spectral form factor, introduced in Ref.~\cite{JiachenLi-21}.
The dissipative spectral form factor captures the complex-spectral correlations of non-Hermitian operators.
By contrast, the dissipative form factor in Eq.~(\ref{eq: DFF definition}) is more directly relevant to the open quantum dynamics since it gives the Loschmidt echo and the decoherence rate, as explained below. 
These two quantities are thus complementary 
to understand the quantum chaos of open systems.
%(\eqref{eq: DFF definition} may be dominated by ``disconnected'' contributions rather than connected ones.) 
While we focus on the dissipative form factor in this work, it should be worthwhile to study the dissipative spectral form factor of the SYK Lindbladians as future work.

Second, 
the dissipative form factor in Eq.~\eqref{eq: DFF definition}
is related to the Loschmidt echo,
%namely, 
the overlap between the initial
and time-evolved states,
\begin{align}
  \mathrm{Tr}_{\mathcal{H}}\,
  \left[
  \rho(0)
  \rho(T_L)
  \right].
\end{align}
%{\bf check what happens if we consider $\mathrm{tr}\,\rho(t)\rho(0)$}
With the operator-state map,
the Loschmidt echo is written as the overlap between
the two pure states in the doubled Hilbert space,
\begin{align}
  \langle \rho(0)|\rho(T_L) \rangle
  =
  \langle \rho(0)|
  e^{ T_L \mathcal{L}}
  |\rho(0) \rangle,
\end{align}
where
$|\rho(0)\rangle \in \mathcal{H}_+\otimes \mathcal{H}_-$
is the state in the doubled Hilbert space $\mathcal{H}_+\otimes \mathcal{H}_-$ mapped from the density matrix $\rho(0)$.
To make a contact with the dissipative form factor,
one %may consider 
needs
to average over the initial states
$|\rho(0)\rangle$.
For example, if we consider
a set of states generated from a reference state
$\rho_0$ by a unitary rotation,
\begin{align}
  \rho_U = U \rho_0 U^{\dag},
\end{align}
and average over the Haar random measure,
we obtain
\begin{align}
  \label{avg over init states}
  &
  \int dU\,
  \langle \rho_U|
  e^{ T_L \mathcal{L}}
  |\rho_U \rangle
    \nonumber \\
  &
  \qquad
  =
  \frac{ \mathrm{Tr}\, (\rho^2_0) - 1/L}
  { L^2 -1}\,
  \mathrm{Tr}\, (e^{T_L \mathcal{L}})
  +
  \frac{L - \mathrm{Tr}\, (\rho^2_0)}{L^2-1},
\end{align}
where $L$
is
the dimensions of the Hilbert space
($L= 2^{N/2}$ for the SYK-type models).
Thus, the Loschmidt echo of open quantum systems is given by the dissipative form factor in Eq.~(\ref{eq: DFF definition}).
Here, if we choose $\rho_0$ to be the fully mixed state $\rho_0 = 1/L$, the first term in the right hand side of Eq.~(\ref{avg over init states}) vanishes.
This is consistent with the fact that the fully mixed state cannot be decohered any longer.
We have to avoid such a special reference state to connect the dissipative form factor with the average Loschmidt echo.

%-- 
Finally,
the dissipative form factor in Eq.~\eqref{eq: DFF definition}
is also related to
the decoherence rate~\cite{Xu-19}
averaged over initial states.
The decoherence rate $D$ quantifies the early-time decay of purity, defined by
\begin{align}
  D
  &=
  \left.
  -
  \frac{2\,\mathrm{Tr}\,
  \left[
  \rho (0) (d\rho(t)/dt)\right]}
  { \mathrm{Tr}\, [\rho(0)^2]}
  \right|_{t=0}.
\end{align}
As before we %can 
average over initial states
$\rho_U$, %and consider
leading to
\begin{align}
  D_{{\rm av}}
%  &=
%    \int dU\,
%    \left.
%    \left[
%    -
%    \frac{2 \mathrm{Tr}\,
%    \left[
%    \rho_U(0) (d\rho_U/dt)\right]}
%    { \mathrm{Tr}\, [\rho_U(0)^2]}
%    \right]
%    \right|_{t=0}
%    \nonumber \\
  &=
    - \frac{
    2}{
    \mathrm{Tr}\, [\rho^2_0]
    }
    \left.
    \frac{d}{dt}
    \int dU\,
     \mathrm{Tr}\,
    \left[
    \rho_U(0) \rho_U(t)\right]
    \right|_{t=0}.
\end{align}
Thus, the average decoherence rate $D_{\rm av}$ is given by the time derivative of the dissipative form factor at $t=0$.
In particular, from Eq.~\eqref{avg over init states},
$D_{{\rm av}}$ %can be 
is 
expressed as
\begin{align}
  D_{{\rm av}}
%  &=
%    \frac{
%    -2}{
%    \mathrm{Tr}\, [\rho^2_0]
%    }
%    \frac{ \mathrm{Tr}\, (\rho^2_0) - 1/L}
%    { L^2 -1}\,
%    \left.
%    \frac{d}{dt}
%    \mathrm{Tr}\, e^{t \mathcal{L}}
%    \right|_{t=0}
%    \nonumber \\
  &=
    -\frac{
    2}{
    \mathrm{Tr}\, [\rho^2_0]
    }
    \frac{ \mathrm{Tr}\, (\rho^2_0) - 1/L}
    { L^2 -1}\,
    \mathrm{Tr}\,
    (\mathcal{L}). \label{eq;AvDecoherenceR}
\end{align}

\section{Nonrandom linear jump operators}
\label{Non-random linear jump operators}

\begin{table*}[t]
\caption{Analogy between
  the two-coupled SYK model
  \cite{2018arXiv180400491M}
  and the SYK Lindbladian
  with the nonrandom linear jump operators
  for the small 
  coupling $\mu$.
  The 
  wormhole in the coupled 
  SYK model
  corresponds to 
  the thermofield double
  (TFD)
  state
  at a certain 
  temperature 
  determined by $\mu$.
  Similarly,
  for the Lindbladian 
  SYK model, 
  the late
  time solution 
  corresponds
  to 
  the infinite 
  temperature TFD state,
  which is 
  the stationary state of
  the Lindbladian.}
\begin{ruledtabular}
\begin{tabular}[t]{cc}
  Two-coupled SYK model & SYK Lindbladian with the nonrandom linear dissipators \\
\hline
 Left/Right system & Bra (+)/Ket ($-$) contour \\
 $H = H^L_{\mathrm{SYK}} + (-1)^{\f{q}{2}} 
 H^R_{\mathrm{SYK}} + i \mu \sum_i  \psi_L^i \psi_R^i$  &  $\mathcal{L} =   -iH_{\mathrm{SYK}}^+ + i  (-1)^\f{q}{2} H_{\mathrm{SYK}}^-    -i \mu \sum_i\psi_+^i\psi_-^i - \mu\f{N}{2} \mathbb{I}$ \\
 Inverse temperature $\beta$ & %Periodicity 
 Time $T_L$ \\
 Partition function $\mathrm{Tr}\, (e^{-\beta H})$ 
 & 
 %``Partition function " 
 Dissipative form factor
 $\mathrm{Tr}\, (e^{T_L \mathcal{L}})$ \\
 Energy $E = \ev{H}$& 
 Lindbladian $\ev{\mathcal{L}}$ 
 (average decoherence rate) \\ 
 Specific heat $C 
 %=- \f{\partial ^2 }{\partial T^2} \log Z 
 = \ev{H^2} -\ev{H}^2 $ & \\
 Energy gap & Decay rate \\
 Black hole & Early time complex solution 
 \\
 Wormhole (TFD) 
% ($\ket{G(\mu)} \approx
% | {\rm TFD} (\tilde{\beta}(\mu)) \rangle$) 
% \ket{TFD(\tilde{\beta}(\mu))}$) 
&
Late time real solution 
(infinite temperature TFD)  \\
 Hawking-Page transition & Late time first-order transition
 (real-complex spectral transition?) \\
 N/A & Early time second-order order transition \\
 Real time physics (e.g. chaos exponents) & ? \\ 
\end{tabular}
\end{ruledtabular}
\label{table:MQandNonRandom} 
\end{table*}

%\textcolor{red}{Can we also compare random dissipation SYK models with susy SYK/Wishert SYK or othe models?
%Some Lindbrad counterpart of susy, complex CFT, symmetry breaking and so on?}

%In this section, 
We consider the SYK model with the nonrandom linear jump operators in Eq.~\eqref{eq: non-random jumps}.
%The first thing to notice here is 
We notice that this open quantum model resembles the two-coupled SYK model  
(Maldacena-Qi model)~\cite{2018arXiv180400491M},
although the SYK Lindbladian is non-Hermitian
while the two-coupled SYK model is Hermitian.
As we will show below, 
there are many analogies %we can make 
between 
these models.  
In Table~\ref{table:MQandNonRandom},
we summarize the similarities and differences between the two-coupled SYK model 
and the SYK Lindbladian with the linear jump operators.
In fact,
one obtains the SYK Lindbladian from the two-coupled SYK model by 
%a simple 
an 
analytical continuation
of the coupling $\mu$ from a real to pure imaginary value. %,at least superficially. 
%looks like an analytic continuation of $\mu$
%to the imaginary direction.
%from the two-coupled SYK model.
In the Hermitian two-coupled SYK model,
the finite temperature partition function 
was studied and the Hawking-Page transition 
(the thermal phase transition between
the black hole and wormhole phases) 
was identified 
\cite{2018arXiv180400491M}.
On the other hand, 
we are here interested in the dissipative form factor in Eq.~\eqref{eq: DFF definition},
where $T_L$ plays the role of the real (rather than imaginary) time.
%We also note that  while in the hermitian two-coupled SYK model we can study the finite temperature partition function, we are interested in the dissipative form factor \eqref{eq: DFF definition}, where $T_L$ is the real (rather than imaginary) time. {\color{magenta} AK: This sentence is a bit unclear. In principle, real time ``partition function" can also be studied for the hermitian case, right?}
%\subsection{Comparison to Maldacena-Qi model }
%The non-Random Lindbrad SYK model has similar property with the Maldacema-Qi model (two coupled SYK), so we summarize here the quantities  we can compare in table \ref{table:MQandNonRandom}. 
%The details will be present in the following. 

\begin{figure*}[t]
\centering

\begin{subfigure}{0.45 \textwidth}
\subcaption{$\mu=0.1$, large $N$}
\includegraphics[scale=0.45]{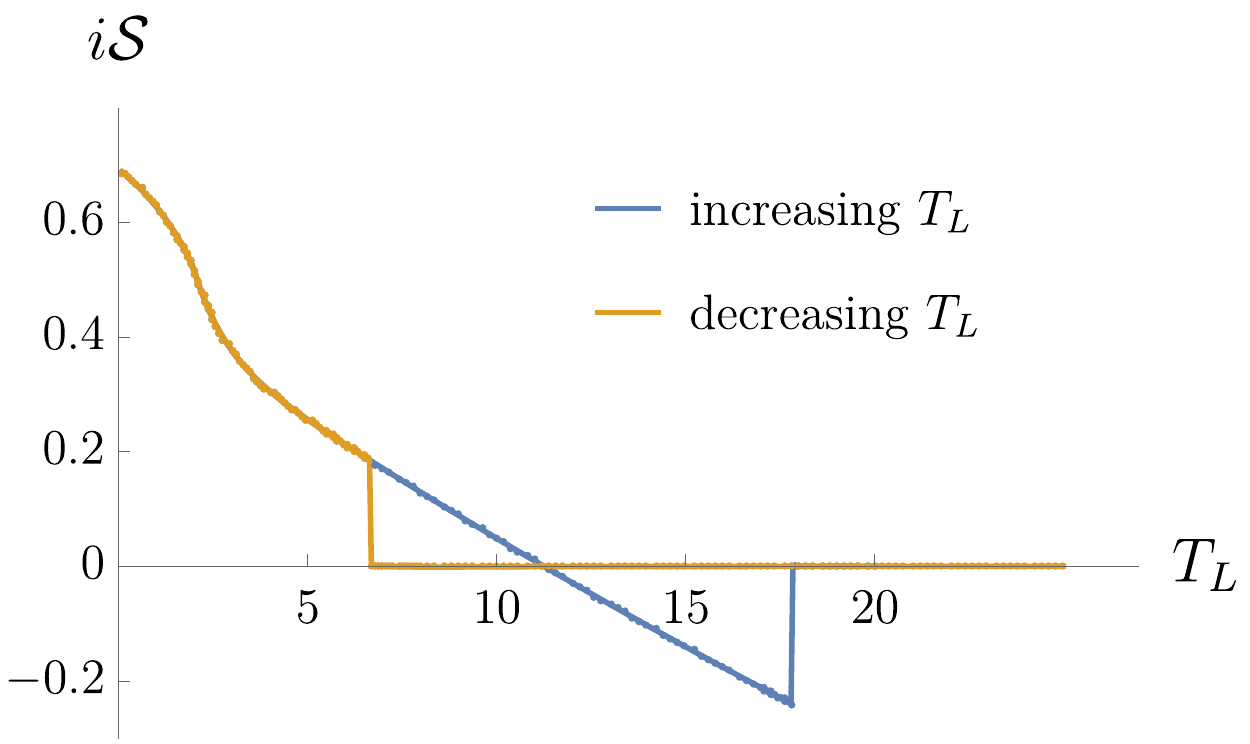}
\end{subfigure}
\begin{subfigure}{0.45\textwidth}
\subcaption{$\mu=0.5$, large $N$}
\includegraphics[scale=0.45]{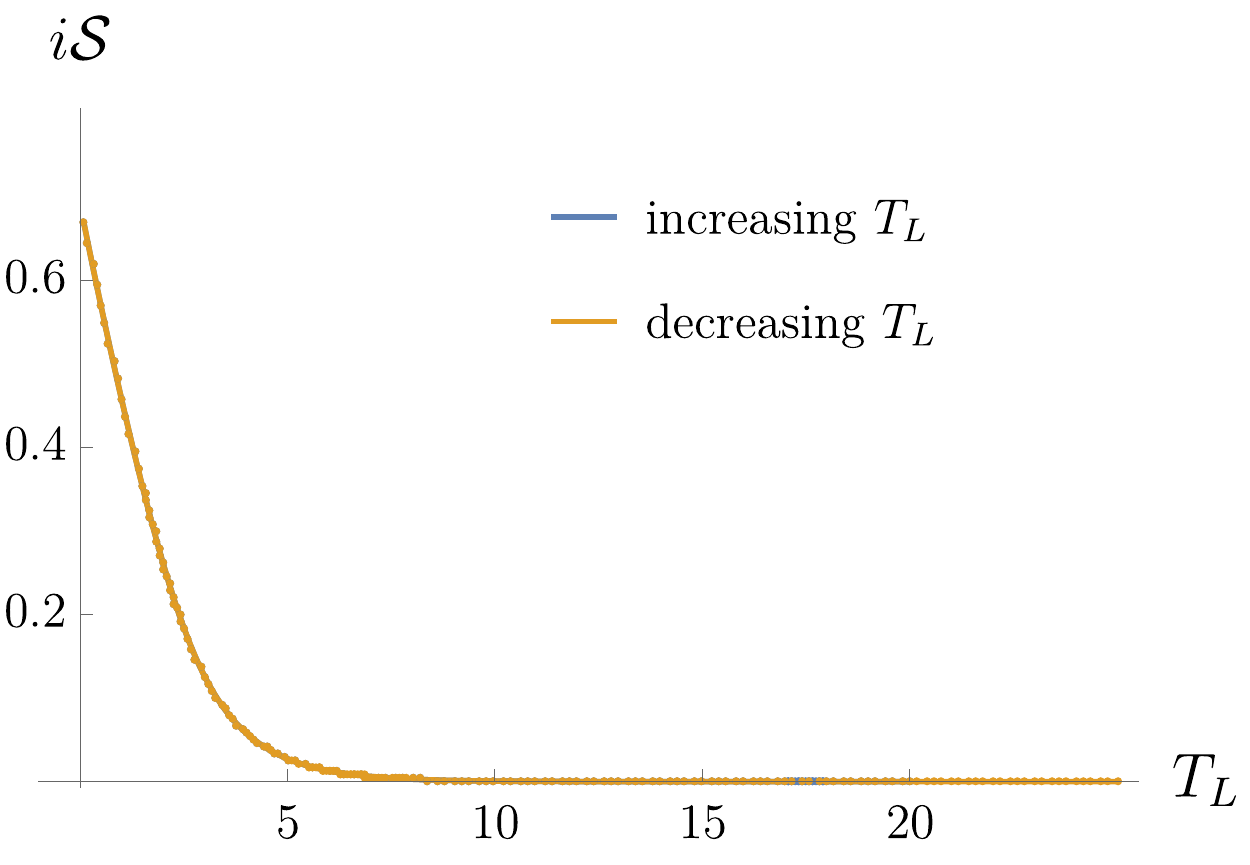}
\end{subfigure}

\begin{subfigure}{0.45 \textwidth}
\subcaption{$\mu=0.1$, large $N$}
\includegraphics[scale=0.45]{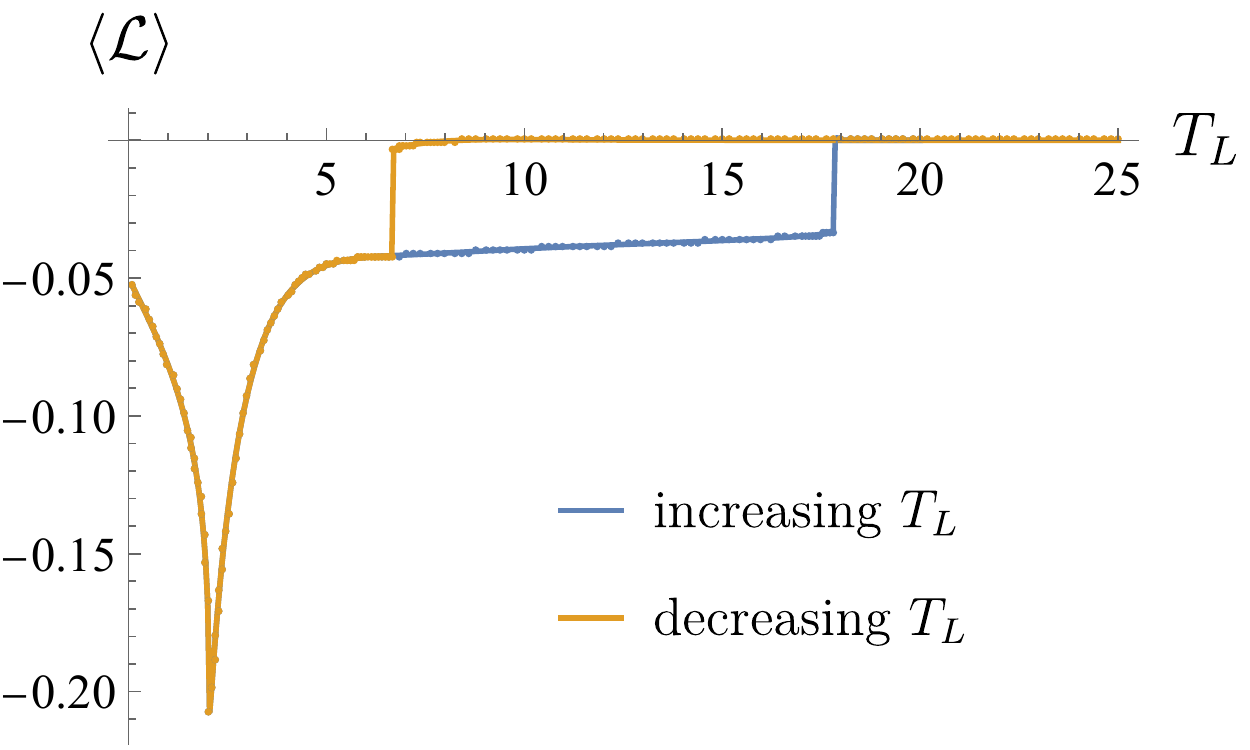}
\end{subfigure}
\begin{subfigure}{0.45\textwidth}
\subcaption{$\mu=0.5$, large $N$}
\includegraphics[scale=0.45]{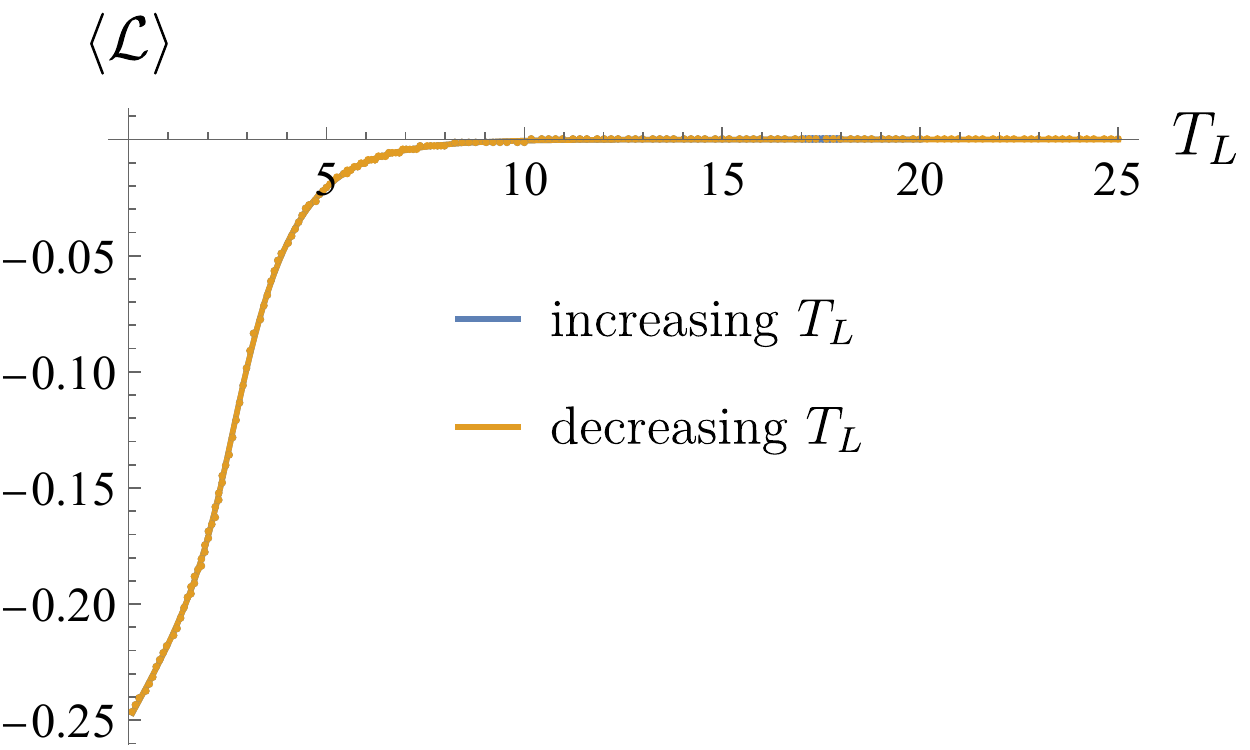}
\end{subfigure}

\begin{subfigure}{0.45 \textwidth}
\subcaption{$\mu=0.1$, finite $N$}
\includegraphics[scale=0.45]{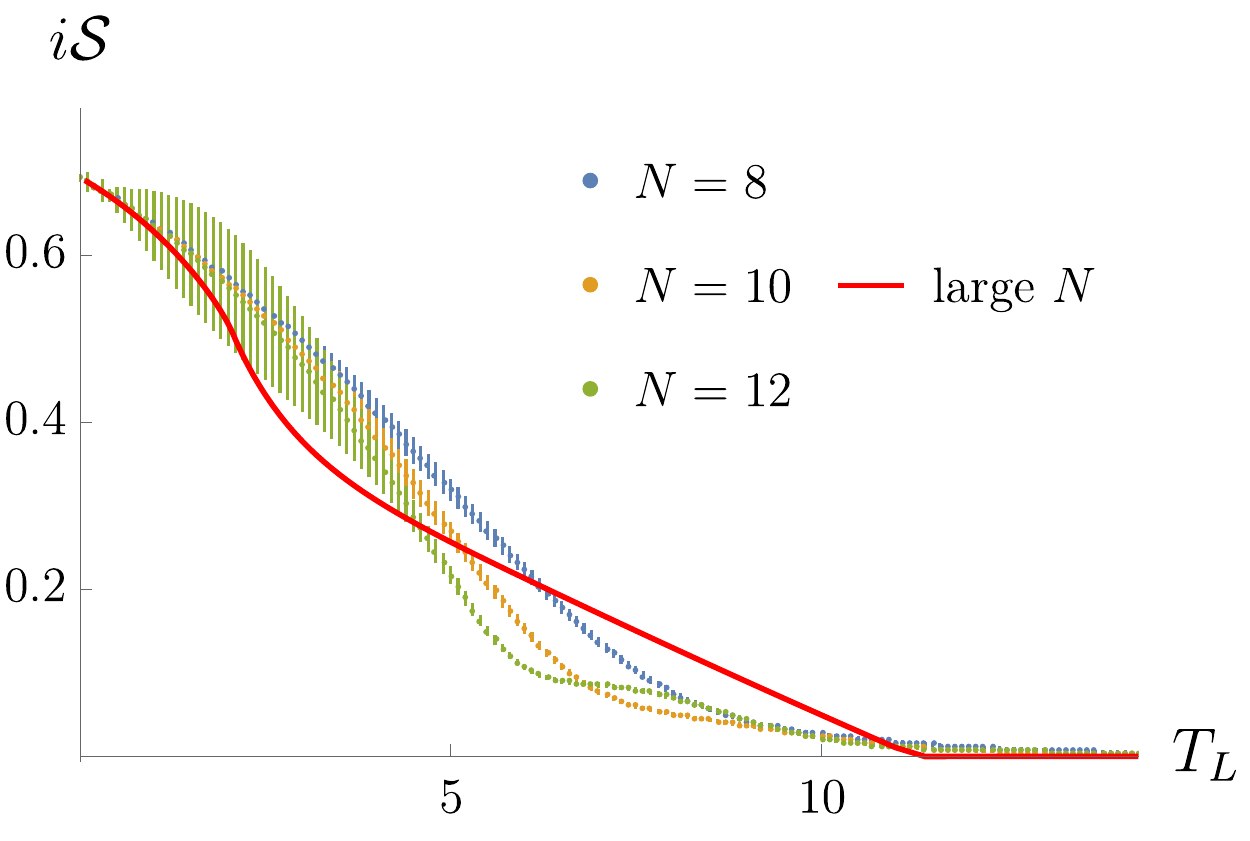}
\end{subfigure}
\begin{subfigure}{0.45\textwidth}
\subcaption{$\mu=0.5$, finite $N$}
\includegraphics[scale=0.45]{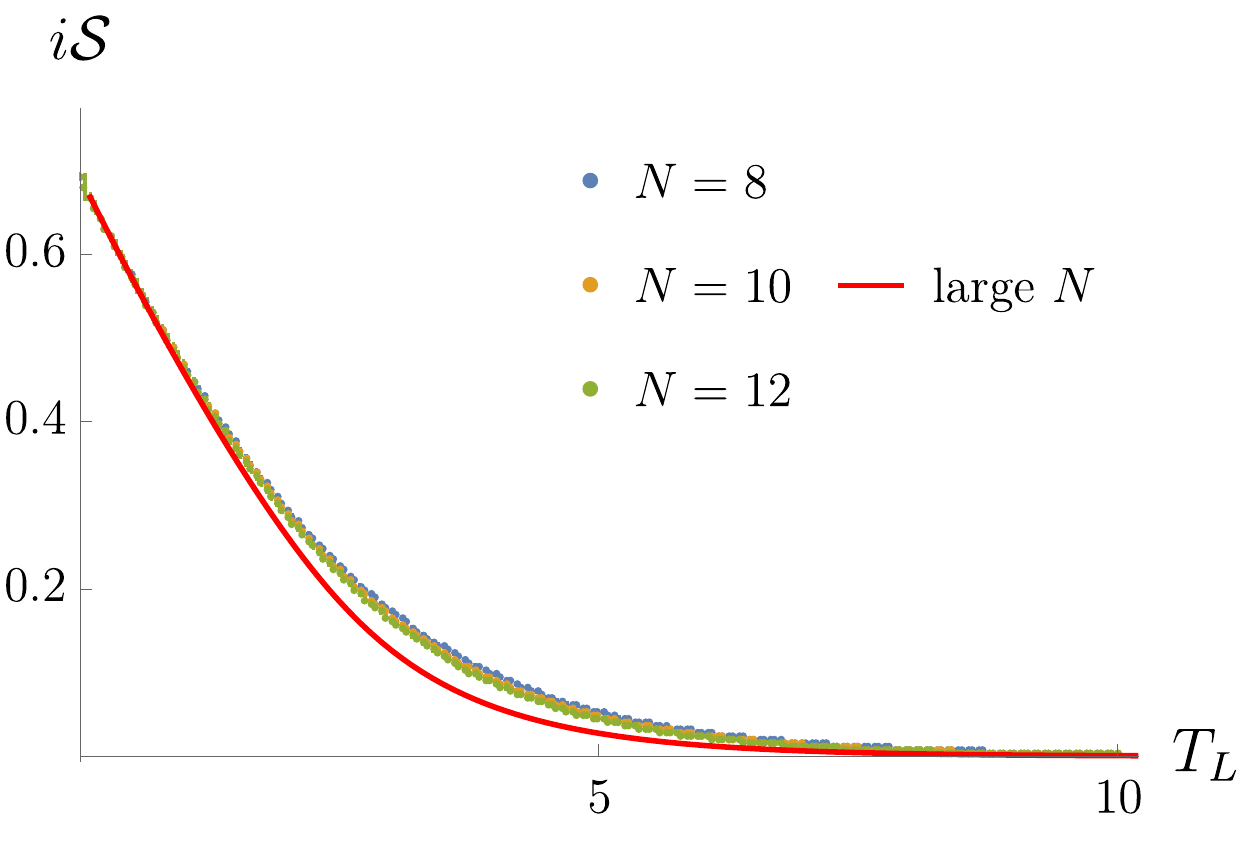}
\end{subfigure}
%  \includegraphics[scale=0.7]{fig1.pdf}
%  \\
%  \includegraphics[scale=0.15]{fig2.png}
%  \includegraphics[scale=0.15]{fig3.png}
%  \includegraphics[scale=0.9]{image775.png}
%\centering
%\begin{subfigure}{0.45 \textwidth}
%\subcaption{$\mu=0.1$, large $N$}
%\includegraphics[scale=0.45]{lexpm01.pdf}
%%\end{subfigure}
%\begin{subfigure}{0.45\textwidth}
%\subcaption{$\mu=0.5$, large $N$}
%\includegraphics[scale=0.45]{lexpm05.pdf}
%\end{subfigure}
\caption{
\label{loschmidt amp, non-random, linear}
%The Loschmidt amplitudes for $J=1$ and $\mu = 0.1$ and $\mu=0.5$.
Time evolution of the rate function (Loschmidt amplitude) $i \mathcal{S}$ of the SYK Lindbladian ($J=1$) with the nonrandom linear dissipators for 
(a)~weak dissipation $\mu = 0.1$ and 
(b)~strong dissipation $\mu = 0.5$.
We obtain the rate function by solving the large $N$ saddle 
point equation numerically
%and (c, d)~numerical calculations for \textcolor{magenta}{$N=$}. 
%In the large $N$ analysis, we obtain the saddle point solution 
by increasing (blue curve) and decreasing (orange curve) $T_{L}$. 
(a)~In the weak-dissipation regime, the two saddle point solutions are different from each other for $7 \lessapprox T_{L} \lessapprox 18$, which signals a discontinuous phase transition around $T_{L} \approx 11$. The second-order derivative of the rate function $i \mathcal{S}$ is discontinuous around $T_{L} \approx 2$, which corresponds to a continuous phase transition. (b)~In the strong-dissipation regime, no dynamical quantum phase transitions occur.
In (c) and (d), the derivatives of the rate functions are shown, which are equivalent to the expectation values $\langle \mathcal{L} \rangle$ of the Lindbladians [see Eq.~\eqref{eq:rateL}].
In (e) and (f), comparisons with finite $N$ 
results from exact diagonalization of the Lindbladian
are presented for $N=8,10,12$,
where the error bars show 
sample-to-sample fluctuations.}
\end{figure*}

\begin{widetext}
\subsection{Large $N$ analysis}

The steady-state Green's functions
of the SYK Lindbladian were studied
in Ref.~\cite{kulkarni2021syk}
%, S__2022}
on the basis of the large $N$ techniques. 
Here, we study 
the dissipative form factor %can be studied
in a similar fashion, by imposing
the %usual 
anti-periodic boundary conditions
for the fermion fields. 
Here, the anti-periodic boundary conditions
arise naturally from the coherent state path integral
representation of $\mathrm{Tr}\, e^{T_L {\cal L}}$,
in much the same way as the 
regular Euclidean (imaginary time) path integral.
Under these boundary conditions, 
the dissipative form factor is given by
the Schwinger-Keldysh path integral, 
\be
F(T_L) = \int \mathcal{D}\psi_+\mathcal{D}\psi_- e^{iS[\psi_+,\psi_-]},
\ee
where the action $iS[\psi_+,\psi_-]$ is 
\ba
iS[\psi_+,\psi_-]  
&= \int_{0}^{T_L} dt \Big[ -\f{1}{2} \sum_i \psi^i_+ \partial_t \psi^i_+-\f{1}{2} \sum_i \psi^i_- \partial_t \psi^i_- + \mathcal{L}(t)\Big] \notag  \\
&= \int_{0}^{T_L} dt \Bigg[ -\f{1}{2} \sum_i \psi^i_+ \partial_t \psi^i_+-\f{1}{2} \sum_i \psi^i_- \partial_t \psi^i_- 
  - i^{\f{q}{2}+1} \sum_{i_1 <\cdots< i_q} J_{i_1\cdots i_q} \psi_+^{i_1} \cdots \psi_+^{i_q} \notag \\
  &\qquad\qquad 
  - (-i)^{\f{q}{2}+1} \sum_{i_1 <\cdots< i_q} J_{i_1\cdots i_q} \psi_-^{i_1} \cdots \psi_-^{i_q} -i \mu \sum_i\psi_+^i(t) \psi_-^i(t) - \mu \f{N}{2} \int dt 
 \Bigg].
\ea
In terms of the collective variables
$(G,\Sigma)$, 
we rewrite the dissipative form factor as 
\be
F(T_L) = \int \mathcal{D}G\mathcal{D}\Sigma\  e^{iS[G,\Sigma]} \label{eq:GSigmaPInonR},
\ee
where the action $S[G,\Sigma]$ is 
\ba
 S[G,\Sigma]
&= -\f{i N}{2} \Tr \log [-i (G_0^{-1} - \Sigma) ]  + \f{i^{q+1}J^2 N}{2 q} \int_{0}^{T_L} dt_1 dt_2  \sum_{\alpha \beta}s_{\alpha \beta}G_{\alpha\beta}(t_1,t_2)^{q} \notag \\
&\qquad + \f{iN}{2}\int_{0}^{T_L} dt_1 dt_2  \sum_{\alpha \beta} \Sigma_{\alpha\beta}(t_1,t_2) G_{\alpha \beta}(t_1,t_2) - i \f{\mu N}{2} \int_{0}^{T_L} dt [G_{+-}(t,t) - G_{-+}(t,t)] +i \f{\mu N}{2} \int dt.
\ea
Here, $\alpha,\beta = +, -$ and $s_{\alpha\beta}$ is 
\be
s_{++} = s_{--} = 1, \qquad s_{+-} = s_{-+} = -(-1)^{\f{q}{2}}.
\ee
For large $N$, the path integral in Eq.~\eqref{eq:GSigmaPInonR} is dominated by the saddle point.
The large $N$ saddle point equation is 
\ba
&i\partial_{t_1} G_{\alpha \beta}(t_1,t_2) - \int dt_3 \sum_{\gamma = +, -}\Sigma_{\alpha\gamma}(t_1,t_3) G_{\gamma \beta}(t_3,t_2) = \delta _{\alpha\beta}\delta(t_1-t_2), \notag \\
&\Sigma_{\alpha\beta}(t_1,t_2) = -i^{q}J^2 s_{\alpha\beta}G_{\alpha\beta}(t_1,t_2)^{q-1} + \mu\epsilon_{\alpha\beta} \delta(t_1-t_2).  \label{eq:KBequationNonRandom}
\ea
The dissipative form factor is then calculated from the on-shell action:
\be
F(T_L) \approx e^{iS[G_*,\Sigma_*]},
\ee
where $G_*$ and $\Sigma_*$ are solutions 
to Eq.~\eqref{eq:KBequationNonRandom}.

%By taking the derivative of the dissipative form factor, we can consider the expectation value of the Lindbladian:
We compute the expectation value of the Lindbladian from the derivative 
of the dissipative form factor,
%\footnote{
%Here, we note that
%in our problem
%the dissipative form factor
%is expected to be self-averaging.
%}, 
\be
 \f{\partial }{\partial T_L} \log F(T_L)  = \langle \mathcal{L} \rangle  
 \equiv
   \f{
   \overline{
   \Tr (\mathcal{L} e^{T_L\mathcal{L}})}}
   {
   \overline{
   \Tr (e^{T_L\mathcal{L}})}}. \label{eq:rateL}
\ee
In our problem, the dissipative form factor is expected to be self-averaging.
For $T_L = 0$, this quantity essentially gives the average decoherence rate in Eq.~\eqref{eq;AvDecoherenceR}. %and we can think of it as a generalized averaged decoherence rate.
Using the canonical commutation relation, we %can
derive the relation between $\langle\mathcal{L}\rangle$ and the correlation function as follows:
\ba
\f{\langle \mathcal{L} \rangle}{N} 
&= -\f{i}{q}\lim_{t \to 0^+}  \partial_{t} G_{++}(t) -\f{i}{q}\lim_{t \to 0^+}  \partial_{t} G_{--}(t)
+ \mu \Big( 1 - \f{2}{q}\Big) G_{+-}(0) - \f{\mu}{2}.
\ea
Here, the two-point correlation functions depend only on the time difference because of the cyclicity of the trace and concomitant time-translation invariance.
Employing the Kadanoff-Baym equation, we find another representation of $\langle \mathcal{L} \rangle$ as 
\ba
\label{L in terms of G}
\f{\langle \mathcal{L} \rangle}{N}
=  i \f{i^{q+1} J^2}{q}  \int_0^{T_L} \sum_{\alpha,\beta} s_{\alpha\beta} G_{\alpha \beta}(t)^q dt + \mu G_{+-}(0) - \f{\mu}{2}.
\ea
We rewrite the latter representation as 
\be
T_L \f{\partial {\cal S} }{\partial T_L} = J \f{\partial {\cal S} }{\partial J } + \mu \f{\partial {\cal S} }{\partial \mu },  \label{eq:TLJmurelation}
\ee
%{\color{red}Is this $S$ or $\mathcal{S}$?}
which is considered to be an Euler relation for the dissipative form factor.
\end{widetext}

\begin{figure}[tbp]
\centering
%\hspace{0.2cm}
%\includegraphics[width=5.5cm]{Gpmm035.pdf} \\
%\hspace{0.5cm}
%\includegraphics[width=5.5cm]{GppRem035.pdf} \\
%\hspace{0.2cm}
%\includegraphics[width=5.5cm]{lexpm035.pdf}
\includegraphics[width=5.5cm]{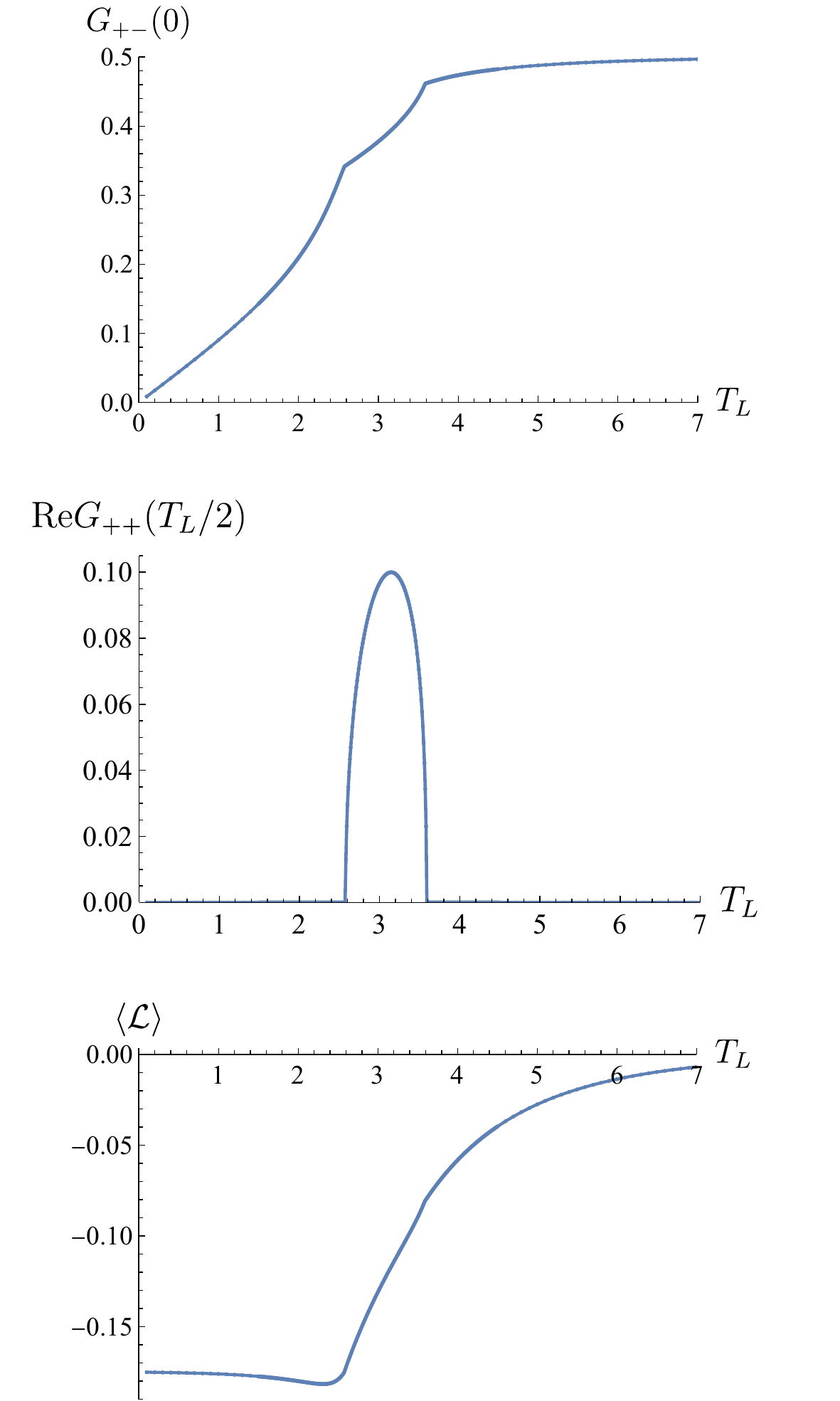}
\caption{ 
%The plots of several quantities as a function of $T_L$ for $q=4, J = 1, \mu = 0.35$ where all the phase transitions are second order.
Second-order dynamical quantum phase transitions in the SYK Lindbladian with the nonrandom linear dissipators ($q=4$, $J = 1$, $\mu = 0.35$).
The Green's functions $G_{+-}(0)$ and $\mathrm{Re}\,G_{++}(T_{L}/2)$ exhibit cusps at $T_{L} \approx 2.6$ and $T_{L} \approx 3.6$, while the singularities in $\langle{\cal L}\rangle$ are unclear.
}
    \label{fig:2ndordermu}
\end{figure}

As shown in Fig.~\ref{loschmidt amp, non-random, linear},
%we plot the rate function $i\mathcal{S}$ for two representative values of $\mu$, $\mu=0.1$ and $\mu=0.5$, at $J=1$.
%At weak dissipation $\mu$,
%we see some interesting features,
%as seen in Fig.\ \ref{loschmidt amp, non-random, linear}.
%Here we have fixed
%$J=1,\mu=0.1$.
%In the figures,
%two different saddle point solutions,
%obtained by increasing/decreasing $T_L$,
%are plotted.
we calculate the rate function $i\mathcal{S}$ as a function of time $T_{L}$.
In these calculations, we obtain the two different saddle point solutions, depending on whether we increase or decrease $T_{\rm L}$. 
For %a 
given $T_L$,
we need to take the %most 
dominant one
(i.e., maximizing $i\mathcal{S}$).
For the strong dissipation $\mu = 0.5$,
the two saddle point solutions are identical [Fig.~\ref{loschmidt amp, non-random, linear}\,(b)]. 
%, as seen in Fig.\ \ref{loschmidt amp, non-random, linear}.
On the other hand,
for the smaller dissipation $\mu = 0.1$,
the two solutions disagree for the intermediate times $7 \lessapprox T_L \lessapprox 18$, and
the rate function $i \mathcal{S}$ exhibits more complex behaviors as a function of $T_L$.
At early time, %$T_L$,
the Green’s functions of the dominant saddle point closely
resemble the dissipation-free unitary solution.
This saddle corresponds to 
the black hole in the two-coupled SYK model. 
Around
$T_L \approx 2$,
%there is a “second-order phase transition” in the sense that the Loschmidt amplitude (the rate function)  has a discontinuous second derivative.
the second-order derivative of the rate function (Loschmidt amplitude) exhibits a discontinuous change, which signals the continuous dynamical quantum phase transition.
In addition, around $T_L \approx 11$,
a discontinuous phase transition occurs 
when the %“dissipative” 
other solution arising from dissipation becomes dominant and remains so for all
subsequent $T_L$.
This saddle corresponds 
to the wormhole in the two-coupled 
SYK model. 
For
$T_L\to \infty$,
this dissipative solution indeed relaxes to the steady-state Green’s function
obtained in Ref.~\cite{kulkarni2021syk},
and corresponds
to the infinite temperature 
thermofield double state.
It is worth recalling that, 
in the finite $N$ spectral analysis
in Ref.~\cite{kulkarni2021syk}, 
all the eigenvalues approach 
%to 
the real axis
as we increase the dissipation strength $\mu$,
similar to a real-complex spectral transition in %some
non-Hermitian systems~\cite{Bender-02, Hamazaki-19}.

\begin{figure}[tbp]
    \centering
    \includegraphics[scale=0.5]{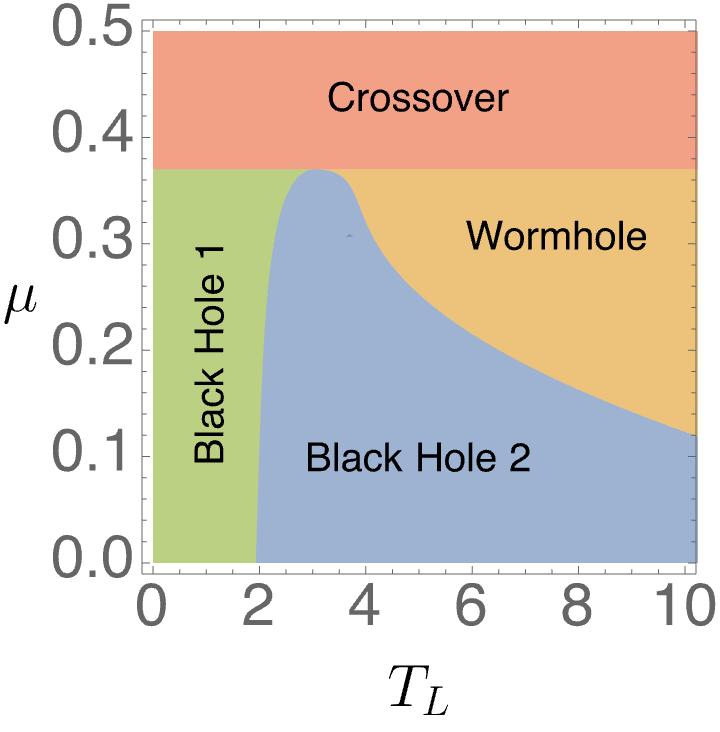}
    \caption{
    Dynamical phase diagram of the SYK Lindbladian with the nonrandom linear jump operators 
    in terms of time $T_L$ and dissipation strength $\mu$ ($q=4$, $J=1$).
    The three phases are defined by $G_{++}(t) \in \mathbb R$, $G_{+-}(t)\not\equiv 0$ (green region, Black Hole 1), $G_{++}(t) \in \mathbb C$, $G_{+-}(t)\not\equiv 0$ (blue region, Black Hole 2), and $G_{++}(t) \in \mathbb R$, $G_{+-}(t) \not\equiv 0$ (orange region, Wormhole).
    For the strong dissipation $\mu \gtrapprox 0.37$, sharp quantum phase transitions are replaced by crossover, 
    and there are no clear distinction between Black Hole 1 phase and Wormhole phase through the 
    order parameters 
    $G_{++}(t)$ and $G_{+-}(t)$.
    }
        \label{fig:phase-diagram-linear-model}
\end{figure}

As we change the dissipation strength $\mu$,
the presence or absence of these transitions,
and also their characters, change.
For small enough $\mu$,
we have the first- and second-order 
phase transitions as described above.
As we increase $\mu$, 
for the intermediate values of $\mu$,
the first-order transition 
is transmuted into second-order,
and
we have two second-order transitions.
In Fig.~\ref{fig:2ndordermu},
we demonstrate the existence of 
the two second-order phase transitions
by plotting the Green's functions
at particular times, $G_{+-}$ at $t=0$ and the real part of $G_{++}$ at $t=T_L/2$, in addition to 
the time derivative of the rate function 
[i.e., expectation value $\langle \mathcal{L} \rangle$ of the Lindbladian; see Eq.~(\ref{eq:rateL})].
While the singularities in 
$\langle{\cal L}\rangle$ are somewhat difficult to see in these plots, the Green's functions exhibit clearer cusp behaviors.
Finally, for large enough $\mu$,
the two second-order transitions
merge and we do not have any transitions.
%\magenta
{We provide the phase diagram in Fig.~\ref{fig:phase-diagram-linear-model}.}
In the next subsection,
we confirm these behaviors 
analytically in the large $q$ limit (see, e.g., Fig.~\ref{fig:TLsigma}).

%These large $N$ results can be compared with finite $N$ numerics. 
Furthermore, we obtain the rate function also for finite $N$ and compare the finite $N$ numerical results with the large $N$ analytical results [Fig.~\ref{loschmidt amp, non-random, linear}\,(c, d)].
%In Fig.\ \ref{loschmidt amp, non-random, linear},
%the Loschmidt amplitude (the rate function)
%computed from finite $N$ numerics 
%is compared with the large $N$ result above.
While %we do not expect the complete agreement, we do see that 
the phase transitions sharply occur only for large $N$, 
the characteristic behaviors of the phase transitions % in large $N$, i.e., 
%the first- and second-order dynamical phase transitions,
have %some 
an
inkling already in
the finite $N$ numerics.
%First, 
In fact,
the %finite $N$ 
rate functions for finite $N$
diminish to nearly zero
around the expected %first-order 
discontinuous phase transition point $T_{L} \approx  11$.
%Second, 
In addition,
%near the expected %second-order 
%continuous phase 
%transition point 
around
$T_{L} \approx 2$, 
the sample-to-sample fluctuations of 
the %finite $N$ 
rate function are enhanced, 
which is consistent with the continuous phase transition for large $N$.

\subsection{Large $q$ analysis}

\begin{figure*}[tbp]
\centering
\includegraphics[width=5.1cm]{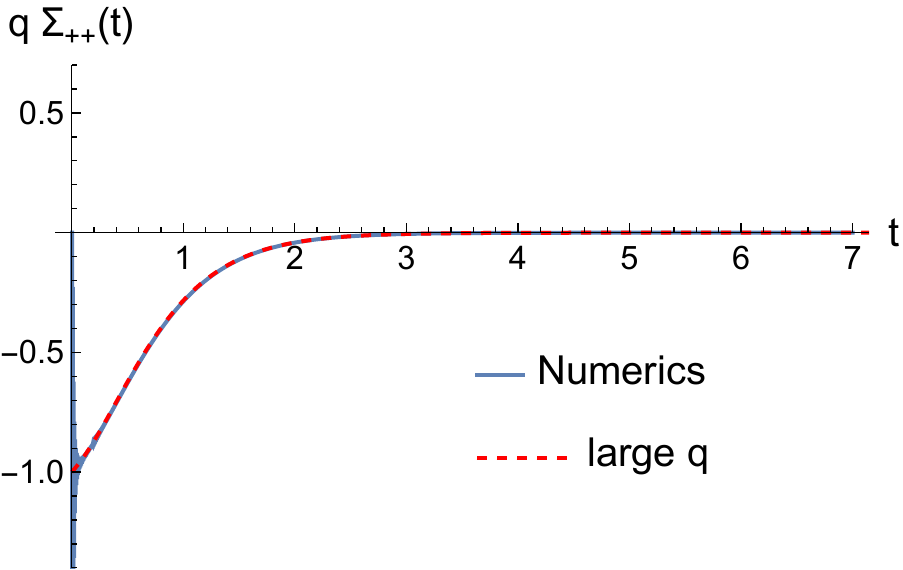}
\hspace{1cm}
\includegraphics[width=5.1cm]{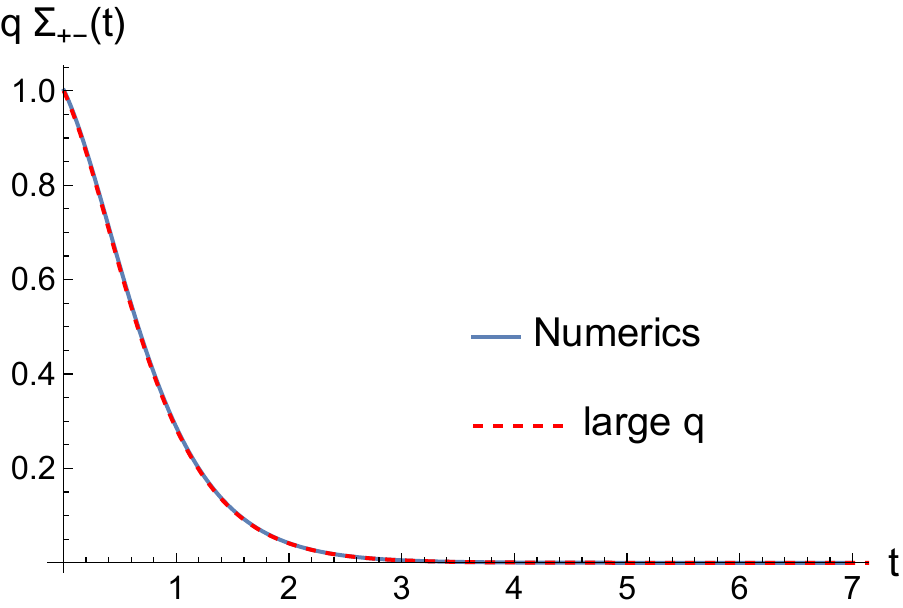}
\caption{ 
Comparison between the analytically obtained real solution for large $q$ (red dashed curve) and the numerically obtained solution for $q = 96$ (blue solid curve).
The other parameters are chosen to be 
%and the numerical solutions ($q=96$, 
$\mathcal{J} = 1$, $\hat{\mu} = 0.5$,
and 
$T_L = 700$. %).
%Here, $q=96$, $\mathcal{J} = 1$, $\hat{\mu} = 3$, and $T_L = 700$. 
The vertical axis is  $q \Sigma_{\alpha \beta} \simeq \mathcal{J}^2 e^{g_{\alpha\beta} }$.
%In the right figure we compare the numerically evaluated  action $iS/N$ for the same $q, \hat{\mu}$  and compare with the large $q$ results in \eqref{eq:largeqActionTLqlogq}.
}
\label{fig:WHcompare}
\end{figure*}

\begin{figure*}[tbp]
\centering
\includegraphics[width=5.1cm]{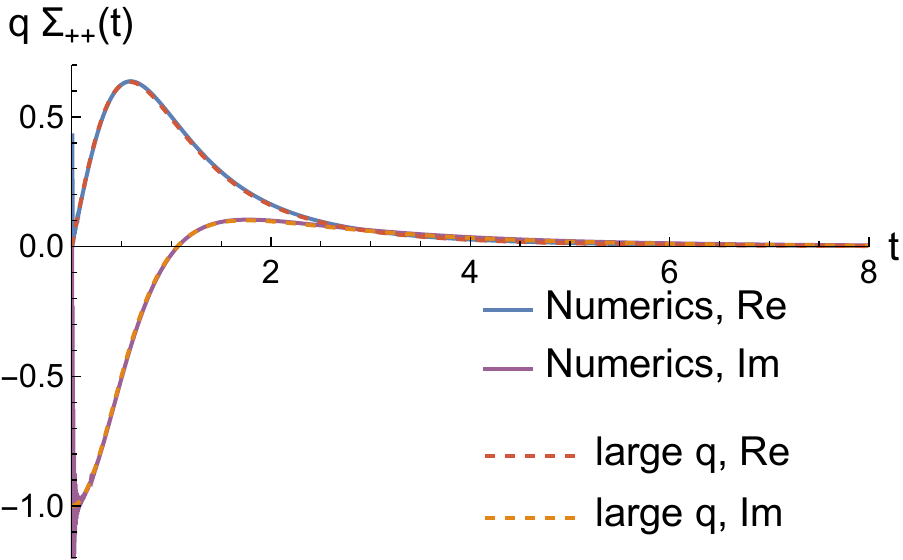}
\hspace{1cm}
\includegraphics[width=5.1cm]{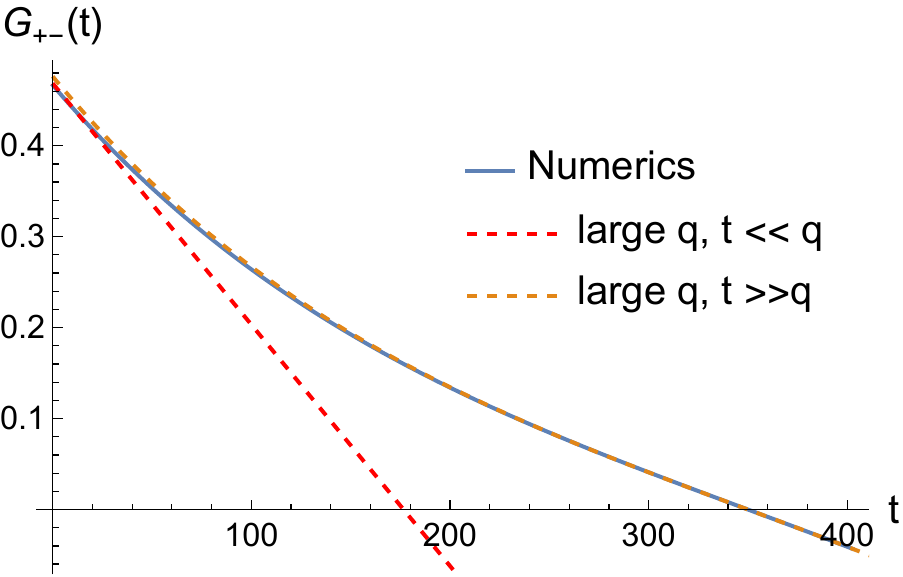}
\caption{
Comparison between the large $q$ complex solution and 
the numerical solutions ($q=96$, $\mathcal{J} = 1$, $\hat{\mu} = 0.5$, $T_L = 700$). 
%The plot is $q \Sigma_{\alpha \beta}$, which is numerically obtained, and $e^{g_{\alpha \beta}(t)}$ in the large $q$ limit.
%The vertical axis is  $q \Sigma_{\alpha \beta} \approx \mathcal{J}^2 e^{g_{\alpha\beta} }$.
%In the right figure we compare the numerically evaluated  action $iS/N$ for the same $q, \hat{\mu}$  and compare with the large $q$ results in \eqref{eq:largeqActionTLqlogq}.
The vertical axis is  $q \Sigma_{++} \simeq \mathcal{J}^2 e^{g_{++} }$ for the left panel and $G_{ +-}$ for the right panel.
 } 
    \label{fig:BHcompare}
\end{figure*}

The large $N$ saddle point equations are analytically tractable
in the large $q$ limit~\cite{Maldacena-Stanford-16}.
Reference~\cite{kulkarni2021syk}
utilized 
the large $q$ limit
of the SYK Lindbladian
with the nonrandom linear jump operators
and calculated
the stationary properties.
Here, we %will 
apply the 
large $q$ technique to 
calculate the dissipative 
form factor.
%as we will show below.
%\paragraph{"Temperature" of order $T_L = q \log q$}
%\textcolor{red}{If I understand correctly,
%Eq. (18) is not quite 
%for $T_L \sim q \log q$,
%so this heading may be inappropriate?}
The large $q$ analysis has to be done 
separately for different time scales. 
In the following, we %will 
mainly focus 
on 
%Here we study 
the regime where $T_L$ is of order $O(q \log q)$,
where, as we show below, 
both first- and second-order 
transitions mentioned above 
occur. 

We start by expanding the correlation
functions for small $t$ (i.e., $t \ll q$) as
%At early time $t \ll q$, we 
%expand the correlation function as 
\ba
G_{++}(t_1,t_2) &= -\f{i}{2} \text{sgn}(t_1-t_2) \Big( 1 + \f{1}{q} g_{++}(t_1,t_2) + \cdots \Big), \notag \\
G_{+-}(t_1,t_2) &=  +\f{1}{2}  \Big( 1 + \f{1}{q} g_{+-}(t_1,t_2) + \cdots \Big), \notag \\
G_{-+}(t_1,t_2) &= -\f{1}{2}  \Big( 1 + \f{1}{q} g_{-+}(t_1,t_2) + \cdots \Big), \notag \\
G_{--}(t_1,t_2) &= -\f{i}{2} \text{sgn}(t_1-t_2) \Big( 1 + \f{1}{q} g_{--}(t_1,t_2) + \cdots \Big). \label{eq:largeqEarly}
\ea
In the large $q$ limit, the Kadanoff-Baym equation reduces to the Liouville equation, 
\ba
\partial_{t_1}\partial_{t_2} g_{++}(t_1,t_2) &= - 2\mathcal{J}^2 e^{g_{++}(t_1,t_2)}, \notag \\
\partial_{t_1}\partial_{t_2} g_{+-}(t_1,t_2) &= - 2\mathcal{J}^2 e^{g_{+-}(t_1,t_2)} -2\hat{\mu}\delta(t_1-t_2),
\ea
where we define $\mathcal{J}$ and $\hat{\mu}$ by
\begin{equation}
    J^2 \equiv \f{2 ^{q-1}\mathcal{J}^2}{ q},\quad
    \mu \equiv \f{\hat{\mu}}{q}.
\end{equation}
%We impose the boundary condition as 
%\ba
%&g_{++}(t,t) = 0, \qquad \lim_{t_2 \to t_1}\partial_{t_1}g_{+-}(t_1,t_2) = - \hat{\mu}, \notag \\
%& g_{++}(t_1,t_2) - g_{+-}(t_1,t_2) \to 0 \qquad \text{as} \ \ t_1 \to \infty  
%\ea
The Liouville equation admits multiple solutions, as 
in the case of finite $q$. 
In the following, we %will 
study two types of solutions with real $g_{++}$ and complex $g_{++}$,
which we call real and complex solutions,
%where $g_{++}$ is real and complex, 
respectively.

\bigskip
\paragraph{Real solutions.}
We %can 
solve the Liouville equation for stationary states as 
\ba
\label{real sol}
e^{g_{++}(t)} &= \f{\alpha^2 }{\mathcal{J}^2 \cosh^2 (\alpha |t| + \gamma )}, \notag 
\\ 
e^{g_{+-}(t)} &= \f{\tilde{\alpha}^2 }{\mathcal{J}^2 \cosh^2 (\tilde{\alpha} |t| + \tilde{\gamma} )}.
\ea
The boundary conditions $G_{++}(0,0) = G_{--}(0,0) = -i/2$ and $\lim_{t_2 \to  t_1}\partial_{t_1}g_{+-}(t_1,t_2) = - \hat{\mu}$ give the relations 
\be
\alpha = \mathcal{J}\cosh \gamma, \quad 2 \tilde{\alpha} \tanh \tilde{\gamma} = \hat{\mu}. \label{eq:t0bdycond}
\ee
%To satisfy the boundary conditions, we impose 
%\be
% \f{\alpha}{\mathcal{J} \cosh \gamma} = 1, \qquad \hat{\mu} = 2\tilde{\alpha} \tanh \tilde{\gamma},\qquad \alpha = \tilde{\alpha}, \qquad \gamma = \tilde{\gamma},.
%f\ee
%By solving these conditions, we obtain
%\be
%\alpha = \tilde{\alpha} = \mathcal{J} \s{\Big(\f{\hat{\mu}}{2\mathcal{J}}\Big)^2 + 1} , \qquad \gamma = \tilde{\gamma} =  \text{arcsinh}\Big( \f{\hat{\mu}}{2\mathcal{J}} \Big)
%\ee

%Correlation functions behaves $G(t) \sim e^{\f{g(t)}{q}}$ and $\f{2 \alpha}{q} $ corresponds to the decay rate.
%This also qualitatively agrees with the $\mu$ dependence for in $q=4$ cases.
For long time $t \gg q$, $g_{\alpha\beta}$ 
becomes of order $q$, and the expansion in Eq.~\eqref{eq:largeqEarly} breaks down.
Hence, for such a long time, we use a different approximation \cite{2018arXiv180400491M, Khramtsov:2020bvs}.
Because $\Sigma_{\alpha\beta}(t)$ varies much more rapidly than $G_{\alpha\beta}$ in the large $q$ limit, we can approximate $\Sigma_{\alpha\beta}(t)$ by delta functions or its derivative.
From the symmetry of the function, we can approximate 
\ba
&\Sigma_{++}(t) = \Sigma_{--}(t) \simeq \rho \delta'(t), \notag \\ 
&\Sigma_{+-}(t) = -\Sigma_{-+}(t) \simeq \nu \delta(t),
\ea
where $\rho$ and $\nu$ are of order $q$.
However, the convolution with  $\Sigma_{++}(t)$ and $\Sigma_{--}(t)$ leads to the derivative of $G_{\alpha\beta}$, which already exists in the Kadanoff-Baym equation and gives the subleading contribution in the large $q$ expansion.
Therefore, we can ignore them, and the equation becomes
\ba
&i\partial_tG_{++} (t) - \nu G_{-+} (t)  = 0, \notag \\
&i\partial_tG_{+-} (t) - \nu G_{--} (t)  = 0, \notag \\
&i\partial_tG_{-+} (t) + \nu G_{++} (t)  = 0, \notag \\
&i\partial_tG_{--} (t) + \nu G_{+-} (t)  = 0,\label{eq:largeqlargetime}
\ea
where $\nu$ is related to the parameter of the $t\ll  q $ solution by
%\ba
%\nu = \int_{-\infty}^{\infty}\Sigma_{+-}(t) dt &= \int_{-\infty}^{\infty} \f{\mathcal{J}^2}{q} e^{g_{+-}(t)}dt + \mu. \notag \\
%&= \f{2  \tilde{\alpha} }{q} + \mu = \f{\mu}{\tanh \tilde{\gamma}} + \mu.
%\ea 
\be
\nu = \int_{-\infty}^{\infty}\Sigma_{+-}(t) dt =  \f{\mu}{\tanh \tilde{\gamma}}. \label{eq:decayratecondition}
\ee
Here, the delta function $\mu\delta(t)$ is included as the boundary conditions of $g_{+-}(t)$ at the origin by $\partial_t g_{+-}(0) = - \hat{\mu}$.
We %can 
rewrite the above condition as 
\be
T_L \mu = \tanh \tilde{\gamma} \log \f{q}{\sigma}, \label{eq:TsigmaWH}
\ee
where we introduce an order $O(1)$ parameter 
\be
\sigma = q e^{- \nu T_L}.
\ee
%and 
%\be
%\hat{\mu} = 2\tilde{\alpha} \tanh{\tilde{\gamma}} = 2\mathcal{J} \cosh \gamma \tanh{\tilde{\gamma}}.
%\ee
The solution of Eq.~\eqref{eq:largeqlargetime} with the correct  boundary conditions is then
\ba
G_{++}(t) &= -iA \cosh \nu\Big( \f{T_L}{2} - t\Big),
\notag \\
G_{+-}(t) &= A \sinh \nu\Big( \f{T_L}{2} - t\Big),
\ea
with $G_{-+}(t) = -G_{+-}(t)$, $iG_{--}(t) = (iG_{++}(t))^*$, $G_{++}(T_L-t) =  G_{++}(t)$, and $G_{+-}(T_L-t) =  -G_{+-}(t)$.

\begin{figure*}[tbp]
\centering
\includegraphics[width=5.2cm]{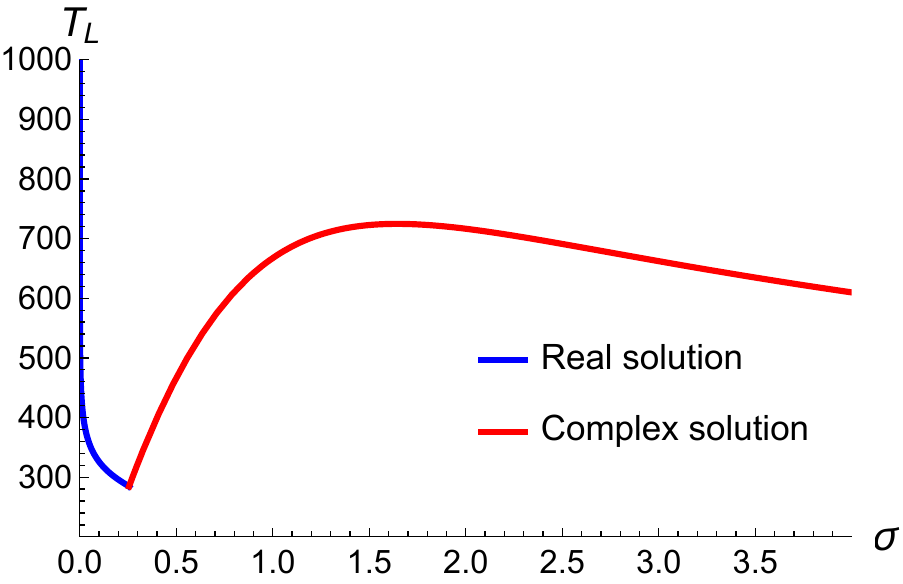}
\hspace{1cm}
\includegraphics[width=5.2cm]{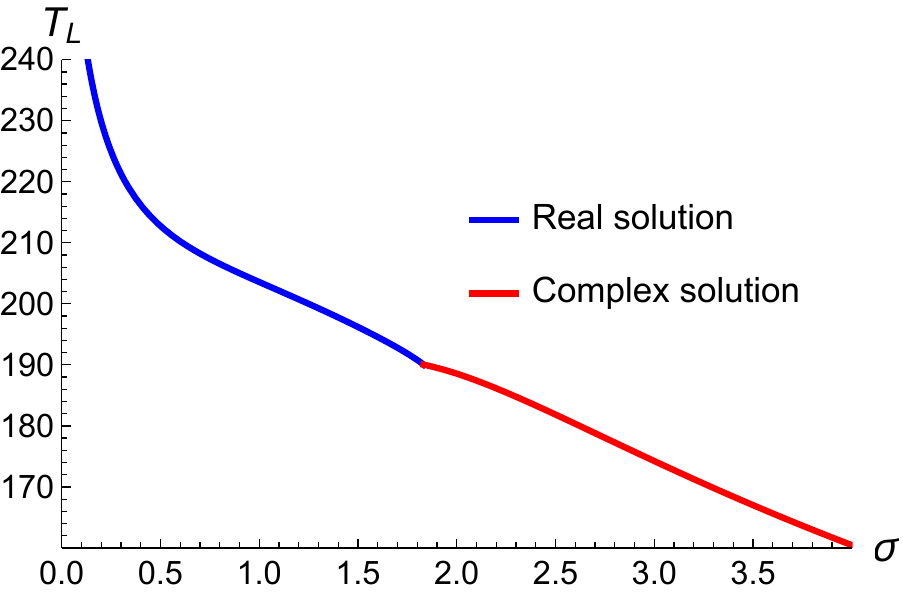}
\caption{$T_L$ as a function of $\sigma $ in 
the large $q$ solutions for $q = 96$.
(Left)~For the weak dissipation $\hat{\mu} = 0.5$,  $T_L$ is not a monotonic function of $\sigma$, and for fixed $T_L$,
there are one real and two complex solutions in the intermediate regime.
(Right)~For the strong dissipation $\hat{\mu} = 1.9$, $T_L$ is a monotonic function of $\sigma$, but the derivative is not continuous when 
the real and complex saddles meet. }
    \label{fig:TLsigma}
\end{figure*}

\begin{figure*}[tbp]
\centering
\includegraphics[width=6.5cm]{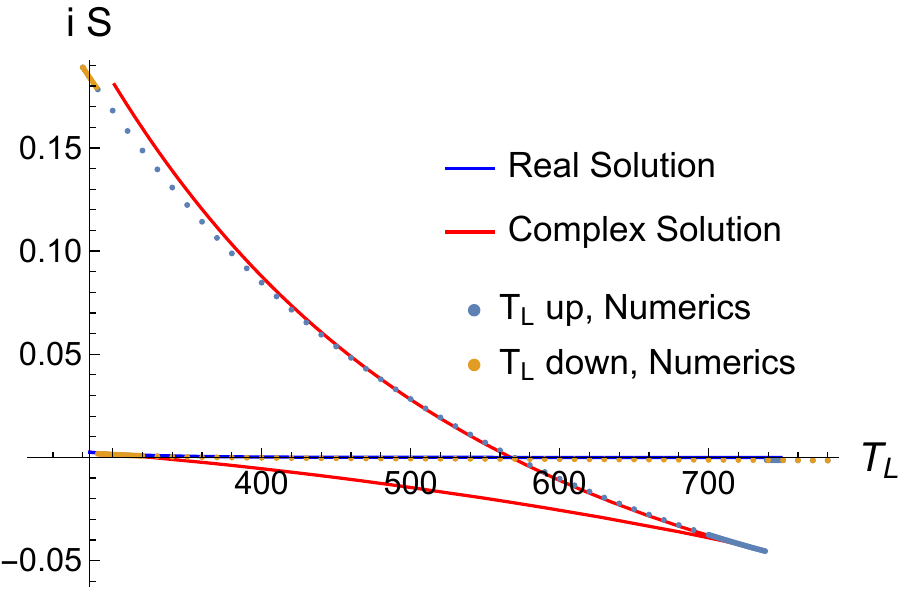}
\includegraphics[width=6.5cm]{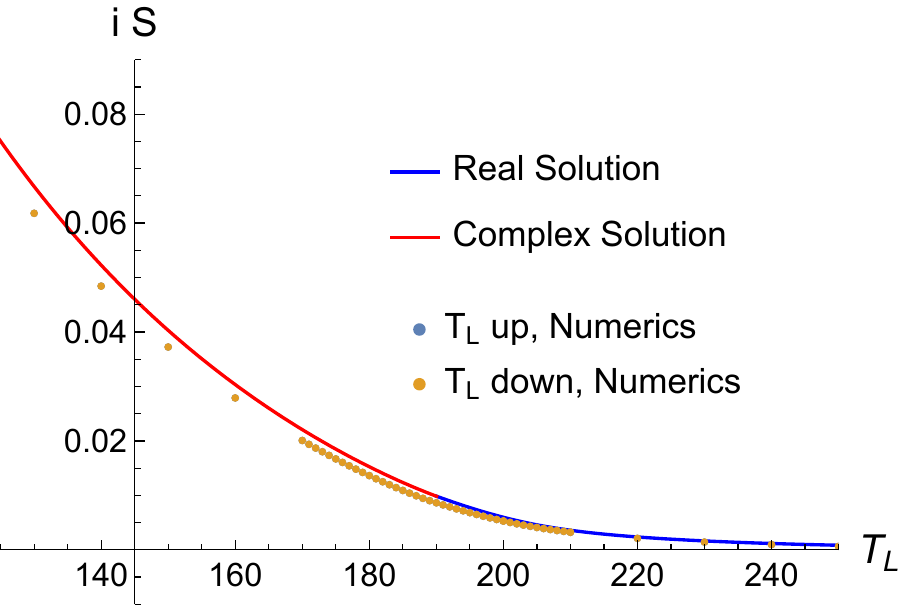}
\caption{
%The on shell action 
Rate function
$i\mathcal{S}$ in the large $q$ limit with $q = 96$ %($\mathcal{J} = 1$) and $\hat{\mu} = 0.5$).
and $\mathcal{J} = 1$
%\magenta
{for (Left)~$\hat{\mu} = 0.5$ and (Right)~~$\hat{\mu} = 1.9$.}
For comparison, 
we also plot the numerical solutions 
of the saddle point equations (blue and orange dots).
%We compare the large $q$ solution with the numerical results.
%{\bf Left:} We take $\hat{\mu} = 0.5$. $T_L$ is not a monotonic function of $\sigma$ and there are two complex solutions in the intermediate regime.
%{\bf Right:} We take $\hat{\mu} = 1.9$. $T_L$ is a monotonic function of $\sigma$ but the derivative is not continuous when the saddle changes from real to complex.
 }
    \label{fig:largeqDFF}
\end{figure*}

The matching of the $t \ll q$ and $t \gg q$ solutions at the overlapping region fixes the parameters as %follow
\be
A = e^{-\f{\nu}{2} T_L}, \quad \alpha = \tilde{\alpha} = \f{q\nu}{2} ,\quad \tilde{\gamma} - \gamma = \sigma. \label{eq:matchingcondition}
\ee
The conditions in Eqs.~\eqref{eq:t0bdycond}, \eqref{eq:decayratecondition}, and \eqref{eq:matchingcondition} determine the free parameters $\alpha,\tilde{\alpha}, \gamma, \tilde{\gamma}, A$, and $\nu$ as functions of $\hat{\mu}$ and $T_L$.
Note %that 
%from equation 
%$\gamma$ is $0$ 
$\gamma = 0$
for $\hat{\mu} = 2\mathcal{J} \tanh \sigma$.
For $\hat{\mu} > 2\mathcal{J} \tanh \sigma$, 
there is no real positive solution
for $\gamma$.
%$\gamma$ does not have real positive solutions.
We compare the large $q$ solution with the numerical solutions for $q=96$ in Fig.~\ref{fig:WHcompare}, which are consistent with each other.

From Eq.~\eqref{L in terms of G},
%the Lindbladian
%Liouvillian 
%operator expectation value is 
%computed from the Green's functions $G_{\alpha\beta}$ and given by
we obtain the expectation value of the Lindbladian as
\be
\f{\ev{\mathcal{L}}}{N} = \f{ \hat{\mu}}{q^2} \bigg( \f{\tanh \gamma}{\tanh \tilde{\gamma}} - 1 +\log \f{\cosh \gamma}{\cosh \tilde{\gamma}}   \bigg)  \label{eq:LiouvillianLargeq}.
\ee
Accordingly, 
%the logarithm of the dissipative form factor 
the rate function of the dissipative form factor
%$i {\cal S} \equiv  \ell(\sigma , \gamma)$ 
is
%{\color{red}In (9), to define ${\cal S}$  we already divide by $N$..?}
\be
i \mathcal{S}(\sigma, \gamma) =
%\ell(\sigma,\gamma) =
\f{T_L \hat{\mu}}{q^2} \bigg( -1 + \f{\tanh \gamma}{\tanh\tilde{\gamma}} + \log \f{\cosh \gamma}{\cosh \tilde{\gamma}} + \f{\sigma}{\tanh \tilde{\gamma}} \bigg) + \f{\sigma}{q}.
\ee

\bigskip
\paragraph{Complex solutions.}
In the real solution in Eq.~\eqref{real sol}, 
the parameters 
$\alpha$, $\tilde{\alpha}$,
$\gamma$, and $\tilde{\gamma}$ 
are real. 
We now relax this condition
and look for the complex solution. 
In particular, we look for the solution 
of the form
%We next consider another solution, the complex solution.
%It is given by
%The complex solution becomes
\ba
e^{g_{++}(t)} &= \f{\alpha}{ \mathcal{J}^2\cosh^2 (\alpha |t| + i \gamma_i)}, \notag \\ 
e^{g_{+-}(t)} &= \f{\tilde{\alpha}^2}{ \mathcal{J}^2\cosh^2 (\tilde{\alpha} |t| +  \tilde{\gamma})}.
\ea
In particular, the $++$ component takes the same form  %with that of 
as 
the real time finite temperature SYK correlation functions, 
where the inverse temperature is given by $\beta = 2\gamma_i/(\mathcal{J}\cos\gamma_i)$.
% \footnote{The temperature is given by $\beta = \f{2\gamma_i}{\mathcal{J}\cos\gamma_i}$.}.
The boundary conditions $G_{++}(0,0) = G_{--}(0,0) = -i/2$ and $\lim_{t_2 \to  t_1}\partial_{t_1}g_{+-}(t_1,t_2) = - \hat{\mu}$ give the relations
\be
\alpha = \mathcal{J} \cos \gamma_i,
%\quad
%\mbox{and}
\quad
\hat{\mu} = 2 \tilde{\alpha} \tanh \tilde{\gamma}. \label{eq:t0bdyBH}
\ee 
Again, $\nu$ is related to $\tilde{\gamma}$ through 
\be
q\nu = q \int _{-\infty}^{\infty}  \Sigma_{+-}(t) dt = \f{\hat{\mu}}{\tanh \tilde{\gamma}}.
\ee
We again introduce $\sigma = q e^{-\nu T_L}$.
Matching the solution of $|t| \ll q$ and $|t| \gg q$, we obtain
\be
\alpha = \tilde{\alpha}, \quad \tilde{\gamma} = \sigma,
\ee
which determines $\gamma_i$ as 
\be
\cos \gamma_i = \f{\hat{\mu}}{2\mathcal{J} \tanh \tilde{\gamma}} =  \f{\hat{\mu}}{2\mathcal{J} \tanh \sigma}.
\ee
Therefore, all the parameters are 
%everything is 
expressed in terms of $\sigma$.
Note %that  
$\gamma_i = 0$ for $\hat{\mu} = 2\mathcal{J}\tanh \sigma$.
For $\hat{\mu} < 2\mathcal{J}\tanh \sigma$, 
there is no real positive solution for $\gamma_i$.
%$\gamma_i$ does not have positive real solutions.
Then, $\sigma$ is related to given $T_L$ and $\hat{\mu}$ by
\be
T_L = \f{q \tanh \tilde{\gamma}}{\hat{\mu}} \log \f{q}{\sigma} = \f{q \tanh{\sigma}}{\hat{\mu}} \log \f{q}{\sigma}. 
\label{eq:TsigmaBH}
\ee
In Fig.~\ref{fig:BHcompare},
we compare the %large $q$ 
analytical solutions for large $q$ with the numerical solutions for $q=96$.
%They are consistent with each other.
The large $q$ solution with $t \ll q$ is consistent with the numerical solution at the early time but deviates from it with time.
On the other hand, the large $q$ solution with $t \gg q$ well agrees with the numerical solution with a slight deviation at the early time.
The rate function 
%The partition function $\ell$ 
is now written as 
\ba
&
i \mathcal{S}(\gamma,\sigma)
%\ell(\gamma,\sigma) 
=   \f{1}{q} \tanh \tilde{\gamma} \log \f{q}{\sigma}
\nonumber\\
&\quad 
\times \text{Re}\,\bigg( -1 + i\f{\tan \gamma_i}{\tanh\tilde{\gamma}} + \log \f{\cos \gamma_i}{\cosh \tilde{\gamma}} + \f{\sigma}{\tanh \tilde{\gamma}} \bigg) + \f{\sigma}{q}\notag \\
&= \f{T_L \hat{\mu}}{q^2} \bigg( -1  + \log \f{\cos \gamma_i}{\cosh \tilde{\gamma}} + \f{\sigma}{\tanh \tilde{\gamma}} \bigg) + \f{\sigma}{q}. \label{eq:largeqActionTLqlogqBH}
\ea

%\textcolor{blue}{TN: Should I write the derivation of the solutions and partition functions relations in appendix?}

\bigskip
\paragraph{Dissipative form factor as a function of $T_L$.}

We are now ready to study the behavior of the dissipative form factor of the SYK Lindbladian for large $q$.
While we are interested in
the dissipative form factor as a function of $T_L$,
we can instead vary the parameter $\sigma$ and then determine $i \mathcal{S}(\sigma)$ and $T_L(\sigma)$ as functions of $\sigma$.
We therefore first plot $T_L$ as a function of $\sigma$ %for representative values $\hat{\mu}$
in Fig.~\ref{fig:TLsigma}, using Eqs.~\eqref{eq:TsigmaWH} and \eqref{eq:TsigmaBH}.
%{\color{red} Does this plot come from (25) and (37)?}
For sufficiently small $\hat{\mu}/{\mathcal{J}}$, 
we find that $T_L$ is not a monotonic function and 
have three solutions for fixed $T_L$, one real solution and two complex solutions.
For $\hat{\mu}/(2\mathcal{J}) \simeq 1$, on the other hand, 
$T_L$ is a monotonic function of $\sigma$ but not smooth at $\sigma = \text{Arctanh}(\hat{\mu}/{2\mathcal{J}})$. 
For $\hat{\mu}>2 \mathcal{J}$, we do not have complex solutions, and $T_L$ is a smooth monotonic function of $\sigma$.
%\magenta
{We note that the precise location of the weak and strong $\mu$ regions depends on $q$. 
Since we use $q=4$ in Fig.~\ref{loschmidt amp, non-random, linear} and $q=96$ in Fig.~\ref{fig:TLsigma}, the weak and strong $\mu$ regimes are different.}

Next, we study the dissipative form factor.
Since $i \mathcal{S}(\sigma)$ itself is a smooth monotonic function of $\sigma$, 
the phase transitions in $i\mathcal{S}(T_L)$
are determined by 
the phase transitions 
in $T_L(\sigma)$ as a function of $\sigma$.
Therefore, for small $\hat{\mu}$ where $T_L(\sigma)$ is not monotonic, we have the discontinuous phase transition.
On the other hand, for large enough $\hat{\mu}$, we do not have phase transitions.
In the intermediate regime with $\hat{\mu}/(2\mathcal{J}) \simeq 1$ %but still smaller than $1$, 
and $\hat{\mu}/(2\mathcal{J}) < 1$,
we have a continuous phase transition.

In Fig.~\ref{fig:largeqDFF},
we plot the rate function 
as a function of $T_L$
for each of the three saddle 
points (two complex and one real solutions).
For the early time, one of the
complex solutions is dominant
and corresponds to the wormhole saddle,
while at later times, the real solution
is dominant and corresponds to
the black hole saddle. 
We also compare the analytic results for large $q$ with 
the numerics at for $q = 96$, which show a good agreement.

We also find that 
the complex solution 
and the
real solution
%and   branches 
are actually continuously connected by 
the other complex solution, which is difficult to find in finite $q$ numerics.
This is similar to what 
was found in the two-coupled SYK model in Ref.~\cite{2018arXiv180400491M}.
In their model, we have the first order phase transition between 
the black hole and wormhole phases.
We also have a ``hot wormhole" 
(or ``small black hole" in the context
of the ordinary Hawking-Page transition)
solution in the large $q$ limit, which connects 
the two solutions.
This means that we have a continuous behavior of the entropy as a function of energy in the microcanonical ensemble. 
It is an interesting future problem to understand %what is 
the analog of this statement in our open quantum dynamical phase transition. 
%[TN: I still didn't determine the names for $\mathcal{X} \leftrightarrow$ wormhole phase, $\mathcal{Y} \leftrightarrow$ black hole phase and $\mathcal{Y}'\leftrightarrow$ hot wormhole]. } 

\section{Random quadratic jump operators}
\label{Random quadratic jump operators}

We consider the SYK Lindbladian with the random quadratic jump operators described in Eq.~\eqref{p-body jumps} with $p=2$. The model shows first-order and second-order dynamical quantum phase transitions. 
%\subsection{Large $N$ analysis}
Similar to the %linear model, 
case for the nonrandom linear jump operators,
the dissipative form factor %for the quadratic model
of the SYK Lindbladian with the random quadratic jump operators
is expressed as a path integral on a circle of circumference $T_L$ with the anti-periodic boundary conditions for the fermion fields. 
Repeating the procedure in Ref.~\cite{kulkarni2021syk}, 
we introduce auxiliary fields so that the path integral action becomes linear in the jump operators. 
After disorder averaging, the action is expressed in terms of the Green's functions and self-energies of the fermions as well as the auxiliary fields:
\begin{widetext}
\begin{align}
	\frac 1N iS[G, \Sigma, G^b, \Sigma^b]
	=&\ \frac 12 \log\det(i \left[\mathbf{G}^0(t_1,t_2)^{-1} - \mathbf{\Sigma}(t_1,t_2)\right]) 
	- R \log\det( \left[\mathbf{G}^{b0}(t_1,t_2)^{-1} - \mathbf{\Sigma}^b(t_1,t_2)\right]) \nonumber \\
	&- \frac{J^2}{2q} i^{q} \iint_0^{T_L} dt_1 dt_2\ s_{\alpha\beta}G_{\alpha\beta}(t_1,t_2)^q 
	- \frac 12  \iint_0^{T_L} dt_1 dt_2\  \Sigma_{\alpha\beta}(t_1,t_2) G_{\alpha\beta}(t_1,t_2) \nonumber \\
	& + \frac{R K^2}{2p} \iint_0^{T_L} dt_1 dt_2\ G^b_{\alpha\beta}(t_1,t_2) G_{\alpha\beta}(t_1,t_2)^p
	- R \iint_0^{T_L} dt_1 dt_2\  \Sigma^b_{\alpha\beta}(t_1,t_2) G^b_{\alpha\beta}(t_1,t_2),
\label{eq:action-quadratic-model}
\end{align}
\end{widetext}
where $R \equiv M/N$ is the ratio of the number of jump operators to the number of fermion flavors. 
The dissipative form factor in terms of this collective action is given by
\be
F(T_L) = \int \mathcal{D}G\mathcal{D}\Sigma \mathcal{D}G^bb\mathcal{D}\Sigma^b\  e^{iS[G,\Sigma,G^b,\Sigma^b]}.
\label{eq:GSigmaPIquad}
\ee
We take the large $N$ limit while fixing $q, J, K,$ and $R$. 
In this limit, the saddle point of the action gives the dissipative form factor and thereby the rate function. 
The saddle point equations are the same as in Ref.~\cite{kulkarni2021syk}. 
However, the boundary conditions are different. 
For fermions, we have the anti-periodic boundary conditions:
$
G_{\alpha\beta}(t+T_L) = - G_{\alpha\beta}(t)
$,
$
\Sigma_{\alpha\beta}(t+T_L) = -\Sigma_{\alpha\beta}(t)
$.
For the auxiliary fields, we have the periodic boundary conditions:
$
G_{b\alpha\beta}(t+T_L) = G_{b\alpha\beta}(t)
$,
$
\Sigma_{b\alpha\beta}(t+T_L) = \Sigma_{b\alpha\beta}(t)
$. 
The large $N$ rate function is then simply given by the right hand side of Eq.~\eqref{eq:action-quadratic-model} evaluated using the saddle point Green's functions $G^*,G^{b*}$ and self energies $\Sigma^*, \Sigma^{b*}$:
\begin{align}
    i\mathcal S(T_L) = \frac 1N i S[G^*, \Sigma^*, G^{b*}, \Sigma^{b*}].
    \label{eq:rate-function-quadratic}
\end{align}

\begin{figure}[tbp]
    \centering
    \includegraphics[scale=0.5]{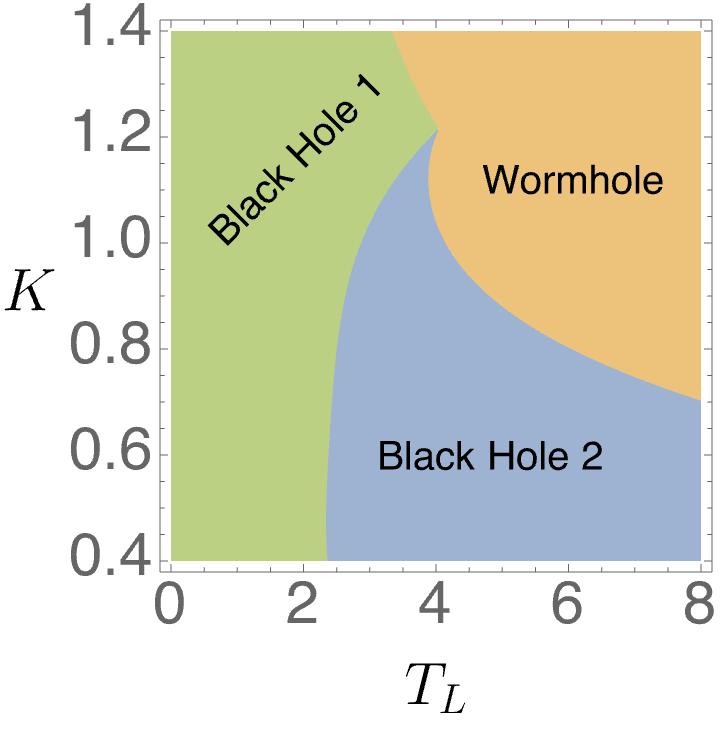}
    \caption{
    Dynamical phase diagram of the SYK Lindbladian with the random quadratic jump operators 
    in terms of time $T_L$ and dissipation strength $K$ ($q=4, J=1$, $R=2$). %. The horizontal axis has $T_L$ and vertical axis has $K$. 
    %We fix $q=4, J=1$ and $R=2$ here. 
    The three phases are defined by $G_{++}(t) \in \mathbb R$, $G_{+-}(t)\equiv 0$ (green region, Black Hole 1), $G_{++}(t) \in \mathbb C$, $G_{+-}(t)\equiv 0$ (blue region, Black Hole 2), and $G_{++}(t) \in \mathbb R$, $G_{+-}(t) \not\equiv 0$ (orange region, Wormhole).
    }
        \label{fig:phase-diagram-quadratic-model}
\end{figure}

\begin{figure*}[tbp]
    \centering
	\begin{subfigure}{0.25 \textwidth} \subcaption{$T_L=2$ (BH1)}
		\includegraphics[width=\textwidth]{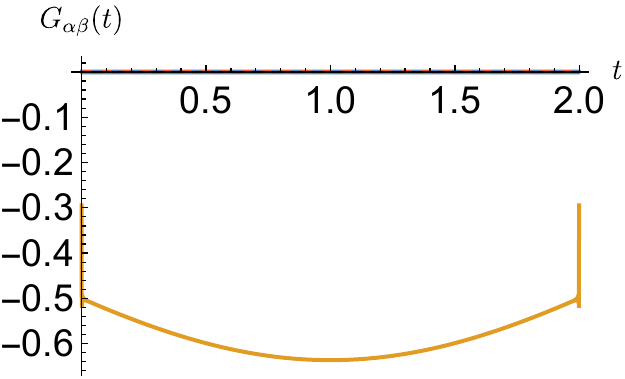}
	\end{subfigure}
	\begin{subfigure}{0.25 \textwidth} \subcaption{$T_L=5$ (BH2)}
		\includegraphics[width=\textwidth]{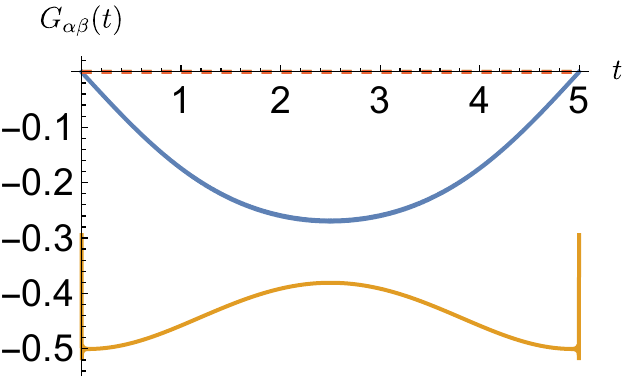}
	\end{subfigure}
	\begin{subfigure}{0.25 \textwidth} \subcaption{$T_L=10$ (WH)}
		\includegraphics[width=\textwidth]{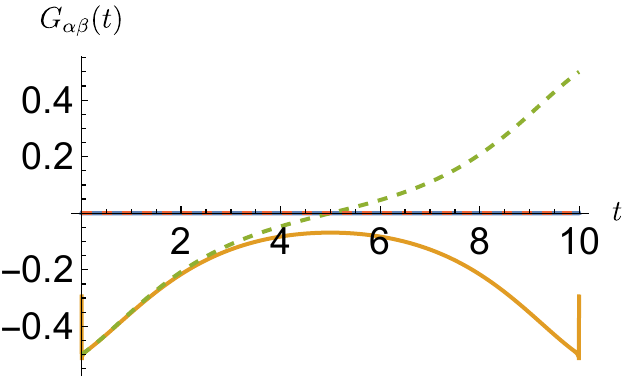}
	\end{subfigure}
	\begin{subfigure}{0.15 \textwidth}
		\includegraphics[width=\textwidth]{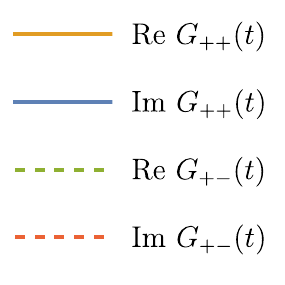}
	\end{subfigure}
	\caption{Representative saddle point Green's functions $G_{++}(t)$ and $G_{+-}(t)$ in the SYK Lindbladian with the quadratic jump operators ($J=1$, $K=0.5$, and $R=2$) for (a)~Black Hole 1 (BH1) phase ($T_{L} = 2$), (b)~Black Hole 2 (BH2) phase ($T_{L}=5$), and (c)~Wormhole (WH) phase ($T_{L} = 10$).}
	\label{fig:quad-model-greens-functions}
\end{figure*}

We solve the saddle point equations numerically and calculate the rate function as a function of $T_L$ for several values of the parameters $J,K,$ and $R$.
We search for various saddle point solutions by slowly increasing (decreasing) $T_L$ starting from a small (large) initial value. 
By doing so, we obtain three kinds of saddle point solutions that correspond to three different dynamical phases
%\magenta
{, summarized as the phase diagram in Fig.~\ref{fig:phase-diagram-quadratic-model}.}

Figure~\ref{fig:quad-model-greens-functions} shows the Green's functions corresponding to these three phases. 
Here, we fix $J=1$, $R=2$, and $K=0.5$, and consider three different values of $T_L$. 
First, Fig.~\ref{fig:quad-model-greens-functions}\,(a) shows the Green's functions at $T_L=2$, characteristic of the early time phase. 
In this phase, the same-sector Green's functions, namely $G_{++}(t)$ and $G_{--}(t)$, are non-zero but purely real. 
The cross-sector Green's functions $G_{-+}(t)$ and $G_{+-}(t)$ are identically zero, indicating that the $+$ and $-$ sectors are still decoupled. 
Next, Fig.~\ref{fig:quad-model-greens-functions}\,(b) shows the Green's functions at $T_L=5$, characteristic of the intermediate time phase. 
The cross-sector Green's functions are still zero, but the same-sector Green's functions acquire a non-zero imaginary part. 
Since both phases are analogous to the black hole in the two-coupled SYK model, hence we label them as Black Hole 1 (BH1) and Black Hole 2 (BH2), respectively.
Finally, Fig.~\ref{fig:quad-model-greens-functions}\,(c) shows the Green's functions for $T_L=10$, characteristic of the late time phase.
Now the cross-sector Green's functions become non-zero, showing that the $+$ and $-$ sectors are now coupled. This phase is thus analogous to the wormhole in the two-coupled SYK model, so we label it as Wormhole (WH).
Interestingly, $G_{++}(t)$ and $G_{--}(t)$ become once again purely real for the wormhole phase. 
As a consistency check, it is worth noting that for very large $T_L$, the Green's functions relax to the steady-state Green's functions discussed in Ref.~\cite{kulkarni2021syk}. 

Next, let us observe how the rate function behaves as we time evolve through the three dynamical phases.
As an aside, for finite dissipation, we expect $i\mathcal{S}(T_L)\rightarrow 0$ as $T_L\rightarrow \infty$ 
if the steady state is unique. 
However, in our numerics, we %see 
find a divergent piece that grows linearly at late times. 
In the subsequent analysis, we subtract this piece and only consider the %``regularized" 
regularized
rate function defined as
\begin{equation}
i\mathcal{S}_\text{reg}(T_L) \equiv i\mathcal S(T_L) - \frac{K^2 R}{24}T_L.
\end{equation}

\begin{figure*}[tbp]
\centering
	\begin{subfigure}{0.32 \textwidth}
		\subcaption{$K=0.3$}
		\includegraphics[width=\textwidth]{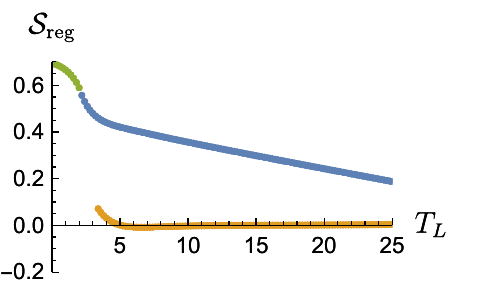}
	\end{subfigure}
	\begin{subfigure}{0.32 \textwidth}
		\subcaption{$K=0.5$}
		\includegraphics[width=\textwidth]{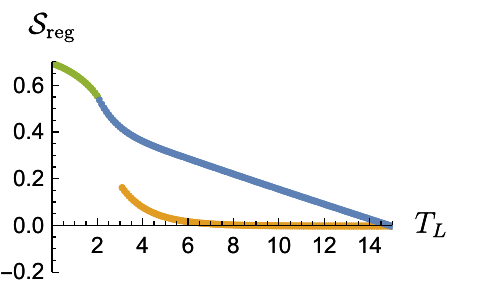}
	\end{subfigure}
	\begin{subfigure}{0.32 \textwidth}
		\subcaption{$K=0.7$}
		\includegraphics[width=\textwidth]{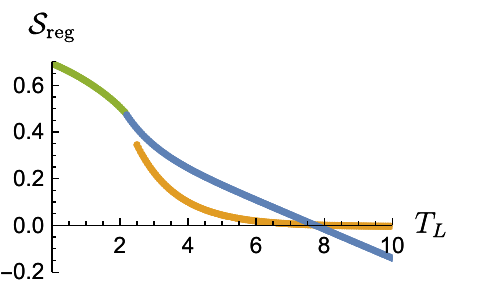}
	\end{subfigure}\\ \vspace{0.5 cm}
	\begin{subfigure}{0.32 \textwidth}
		\subcaption{$K=0.8$}
		\includegraphics[width=\textwidth]{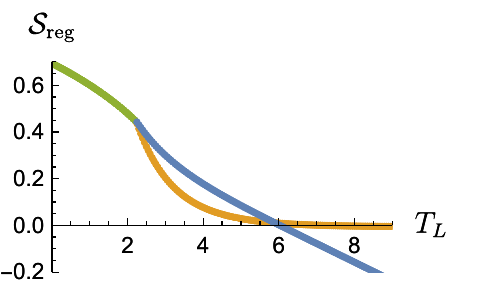}
	\end{subfigure}
	\begin{subfigure}{0.32 \textwidth}
		\subcaption{$K=0.9$}
		\includegraphics[width=\textwidth]{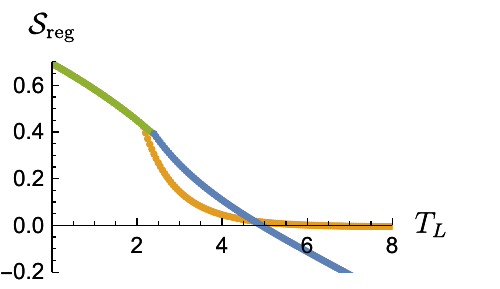}
	\end{subfigure}
	\begin{subfigure}{0.32 \textwidth}
		\subcaption{$K=1$}
		\includegraphics[width=\textwidth]{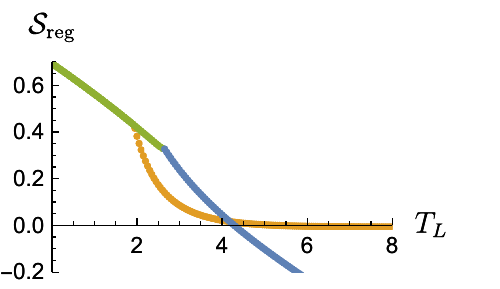}
	\end{subfigure}\\ \vspace{0.5 cm}
	\begin{subfigure}{0.32 \textwidth}
		\subcaption{$K=1.3$}
		\includegraphics[width=\textwidth]{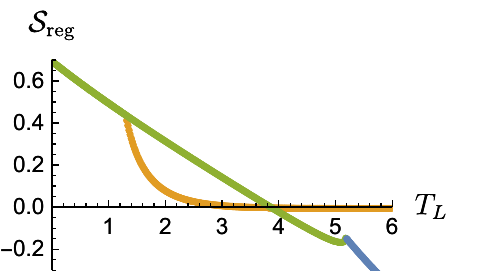}
	\end{subfigure}
	\begin{subfigure}{0.32 \textwidth}
		\subcaption{$K=1.5$}
		\includegraphics[width=\textwidth]{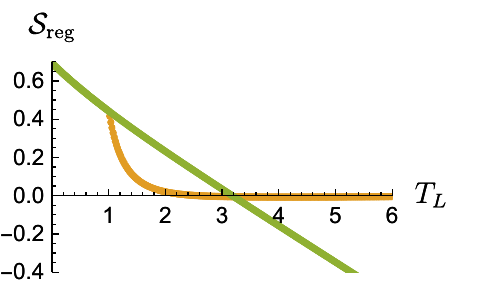}
	\end{subfigure}
	\begin{subfigure}{0.32 \textwidth}
 		\includegraphics[width=\textwidth]{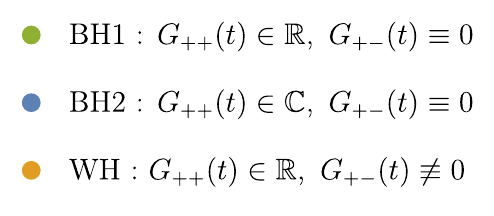}
	\end{subfigure}
	\caption{Time evolution of the rate function in Eq.~(\ref{eq:rate-function-quadratic}) for various values of the dissipative strength $K$ ($q=4, J=1$, $R=2$).}
    \label{fig:quadratic-model-DFF-with-varying-K}
\end{figure*}
%\begin{figure*}[tbp]
%\centering
%	\begin{subfigure}{0.32 \textwidth}
%		\subcaption{$K=0.5$}
%		\includegraphics[width=\textwidth]{partition-func/finite-N/S_R_DHS_diff_N_R2_q4_K0.5_compare.pdf}
%	\end{subfigure}
%	\begin{subfigure}{0.32 \textwidth}
%		\subcaption{$K=1$}
%		\includegraphics[width=\textwidth]{partition-func/finite-N/S_R_DHS_diff_N_R2_q4_K1_compare.pdf}
%	\end{subfigure}
%	\caption{Time evolution of the rate function calculated numerically comparing to that defined in Eq.~(\ref{eq:rate-function-quadratic}) for various values of the dissipative strength $K$. The remaining parameters are $q=4, J=1$ and $R=2$.}
%    \label{fig:quadratic-model-DFF-with-varying-K}
%\end{figure*}

\begin{figure*}[tbp]
	\centering
	\begin{subfigure}{0.3 \textwidth}
		\subcaption{$R=5$}
		\includegraphics[width=\textwidth]{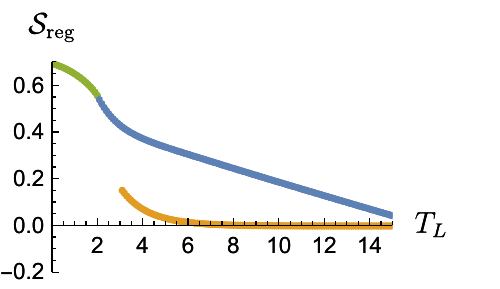}
	\end{subfigure}
	\begin{subfigure}{0.32 \textwidth}
		\subcaption{$R=10$}
		\includegraphics[width=\textwidth]{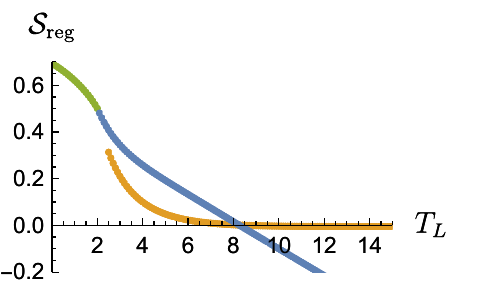}
	\end{subfigure}
	\begin{subfigure}{0.32 \textwidth}
		\subcaption{$R=15$}
		\includegraphics[width=\textwidth]{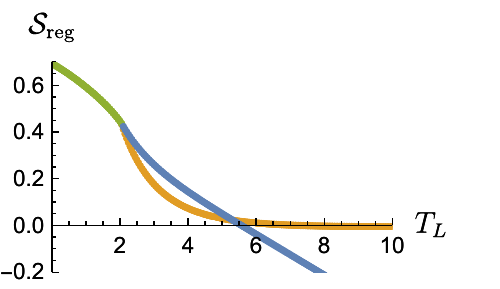}
	\end{subfigure}\\ \vspace{1 cm}
	\begin{subfigure}{0.32 \textwidth}
		\subcaption{$R=20$}
		\includegraphics[width=\textwidth]{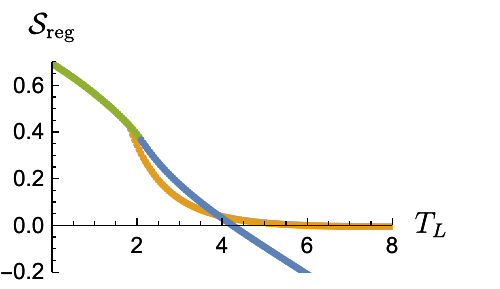}
	\end{subfigure}
	\begin{subfigure}{0.32 \textwidth}
		\subcaption{$R=30$}
		\includegraphics[width=\textwidth]{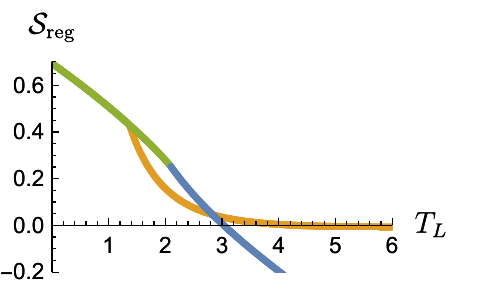}
	\end{subfigure}
	%\hspace*{-6cm}
% 	\begin{subfigure}{0.32 \textwidth}
% 		\includegraphics[width=\textwidth]{images/partition-func/legend.pdf}
% 	\end{subfigure}
	\caption{Time evolution of the rate function in Eq.~(\ref{eq:rate-function-quadratic}) for various values of the dissipative strength $R$ ($q=4, J=1$, $K=0.3$).}
	\label{fig:quadratic-model-DFF-with-varying-R}
\end{figure*}

We first fix $J =1$ and $R = 2$, and vary the dissipation strength $K$ from 0.3 to 1.5. 
The rate function as a function of time $T_L$ is plotted in Fig.~\ref{fig:quadratic-model-DFF-with-varying-K}. 
The legend shows the %color and Green's function
characterization for each phase. 
At every time point, the dominant phase is the one with the larger rate function. 
For $K\lessapprox 1.1$, all the three phases occur. 
The transition from early time (Black Hole 1 phase) to the intermediate time (Black Hole 2 phase) is first order for $K\gtrapprox 0.8$ and second order for $K\lessapprox 0.8$. 
The transition from Black Hole 2 phase to Wormhole phase is always first order. 
For $K\gtrapprox 1$, there are only two phases since Black Hole 2 phase (shown in blue) is always sub-dominant.
We next fix $J=1$ and $K=0.3$, and vary $R$ from 5 to 30. 
The time evolution of the rate function is shown in Fig.~\ref{fig:quadratic-model-DFF-with-varying-R}, which exhibits similar behavior.
%We see very similar behavior: 
We have the three distinct phases even for $R$ as large as $30$. 
The intermediate to late time transition is second order while the early to intermediate time transition is second order for $R\lessapprox 15$ and first order $R\gtrapprox 15$.

The dynamical phase diagram shown in Fig.~\ref{fig:phase-diagram-quadratic-model} summarizes the above results.
%\magenta{
Notably, while the phase diagram for the nonrandom linear dissipators in Fig.~\ref{fig:phase-diagram-linear-model} and that for the random quadratic dissipators in Fig.~\ref{fig:phase-diagram-quadratic-model} look similar to each other, the definitions of the phases differ.
Specifically, the black hole phases in the SYK Lindbladian with the linear dissipators accompany $G_{+-} \neq 0$, which contrasts with those with the quadratic dissipators satisfying $G_{+-} = 0$.
This difference arises from the presence or absence of fermion parity symmetry (in the strong sense~\cite{Albert-14}).
On the one hand, strong fermion parity symmetry is respected in the SYK Lindbladian with the quadratic dissipators and hence can be spontaneously broken.
In fact, the phase transitions between the black hole phases and the wormhole phase can be interpreted as the spontaneous breaking of strong fermion parity symmetry, in which $G_{+-}$ serves as an order parameter.
On the other hand, the SYK Lindbladian with the linear dissipators explicitly breaks strong fermion parity symmetry.
As a result, we generally have $G_{+-} \neq 0$ in all the phases, and the phase transitions cannot be understood as spontaneous symmetry breaking.
%}

%\subsection{Finite $N$ analysis}

\section{Discussion}
\label{Discussion}

In this work, we have investigated %the dissipative form factor for 
the open quantum dynamics of
the SYK Lindbladians %introduced in
and found the dynamical quantum phase transitions of its dissipative form factor.
%As we showed, the dissipative form factor exhibits various
%dynamical phase transitions in the real-time domain.
%While some of these transitions (e.g., in the case of the
%non-random linear jump operators) are somewhat analogous to finite temperature transitions in the coupled SYK
%model, say, we also found new kinds of transitions unique
%to open quantum many-body systems. 
For the nonrandom linear dissipators, 
we have found both discontinuous and continuous dynamical phase transitions.
While the former is formally analogous 
to the thermal phase transition
in the two-coupled SYK model, 
the latter does not have a 
Hermitian counterpart. 
%We have demonstrated the discontinuous dynamical phase transition for the non-random linear dissipators, which is formally analogous to the thermal phase transitions in the coupled SYK model,
%and 
%also the second-order dynamical phase transition that does not have a 
%counter part in the two-coupled SYK model.
For the random quadratic dissipators, 
%on the other hand, 
we have found the continuous phase transition that has no counterparts in the original SYK model.
More precise characterizations of these phase transitions are still lacking and are left as a future problem.

More broadly, there
are many open questions in the far-from-equilibrium
properties of open quantum many-body systems. 
We expect that the SYK Lindbladians studied in this work, and generalizations thereof, can be further investigated
as a prototype of open quantum many-body systems. 
For
example, it would be interesting to calculate the dissipative spectral form factor~\cite{JiachenLi-21, Shivam-22, Ghosh-22},
which may better capture
the complex-spectral correlations of non-Hermitian operators.
It is also notable that the operator growth has recently been studied in generic open quantum systems~\cite{Bhattacharya-22, Liu-Tang-Zhai-22, Schuster-Yao-22}.
Since the SYK Lindbladians should be a prototype for the dissipative quantum chaos, it is worthwhile studying their operator growth.

%-- Maybe we can also mention other quatities, such as OTOC? (CITE: Norm Yao's recent paper?)

\bigskip
{\it Note added.}
While finalizing the manuscript, 
Ref.~\cite{https://doi.org/10.48550/arxiv.2210.01695}
appeared on arXiv,
which has a substantial overlap with the current work.
We also note another related work
\cite{Garc_a_Garc_a_2023} 
that appeared after the submission of 
this work.
%with our section 
%\ref{Many Random Jump Operators (Complex coupling)}.

\section*{Acknowledgments}
This work is supported by MEXT KAKENHI Grant-in-Aid for Transformative Research Areas A ``Extreme Universe'' Grant Number 22H05248,
by the National Science Foundation under 
Award No.~DMR-2001181, and by a Simons Investigator Grant from
the Simons Foundation (Award No.~566116).
This work is supported by the Japan Society for the Promotion of Science (JSPS) through the Overseas Research Fellowship.
This work is supported by
the Gordon and Betty Moore Foundation through Grant
No.~GBMF8685 toward the Princeton theory program.
This work was performed in part at Aspen Center for Physics, which is supported by National Science Foundation grant PHY-1607611. 
This work was partially supported by a grant from the Simons Foundation.

\appendix

\bibliography{ref}% Produces the bibliography via BibTeX.

\end{document}